\documentclass[aps,prl,twocolumn,10pt,superscriptaddress,longbibliography,floatfix,footinbib]{revtex4-2}

\usepackage[T1]{fontenc}
\usepackage[utf8]{inputenc}
\usepackage{amsmath,amsthm,mathtools}
\usepackage{newtxtext}
\usepackage{newtxmath}
\usepackage{graphicx}
\usepackage{xcolor}
\usepackage{microtype}
\usepackage{hyperref}
\hypersetup{colorlinks=true,linkcolor=blue!45!black,citecolor=blue!45!black,urlcolor=blue!45!black,
pdfauthor={Weikang Li},
pdftitle={electrostatic limits on quantum spatial search}}
\newtheorem*{theorem*}{Theorem}
\newcommand{\HH}{\mathcal H}
\newcommand{\Lap}{\mathcal L}
\newcommand{\KK}{\mathsf K}

\newcommand{\Mset}{\mathcal M}
\newcommand{\norm}[1]{\lVert #1\rVert}
\newcommand{\ket}[1]{\lvert #1\rangle}
\newcommand{\bra}[1]{\langle #1\rvert}
\newcommand{\braket}[2]{\langle #1\vert #2\rangle}
\newcommand{\thit}{t_{\mathrm{h}}}
\newcommand{\Tmin}{T_{\min}}
\newcommand{\mainfigureone}{%
\begin{figure*}[!t]
\centering
\includegraphics[width=0.99\textwidth]{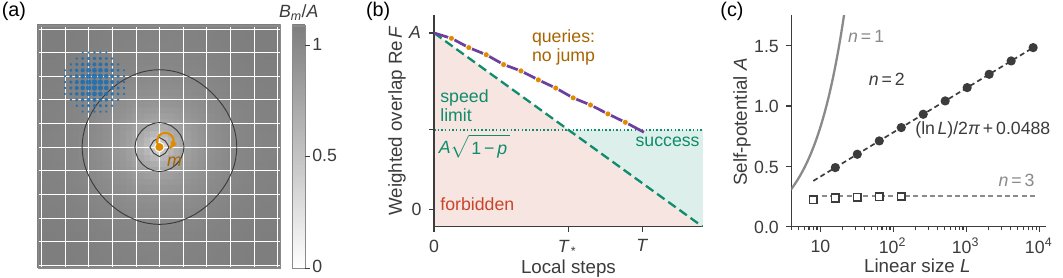}
\caption{%
Electrostatic bound.
(a)~Potential drop $B_m(x)=A-K(x,m)$, the Dirichlet-optimal weight of the overlap, on a $41\times41$ torus in units of $A$, with contours at $B_m/A=0.5$, $0.75$ and $1$ and white lines along every fourth row and column. It vanishes at the marked vertex $m$ (orange), where queries act, and grows logarithmically with distance. Blue dots sketch the walker amplitude.
(b)~Real part of the weighted overlap $F$ for $w=B_m$ along a schematic trajectory (purple). It starts at $A$, does not jump at queries (orange dots) and changes by at most $\sqrt{2A/N}$ per local step [Eq.~\eqref{eq:main-speed}, dashed line], which forbids the red region. Success with probability $p$ requires $\mathrm{Re}\,F\lesssim A\sqrt{1-p}$ (green), up to endpoint terms of order $(A+S)/\sqrt N$, and reaching this region takes at least $T_*$ local steps, the right side of Eq.~\eqref{eq:main-theorem}.
(c)~Self-potential $A$ on the $n$-dimensional torus of linear size $L$. The line is the exact $(L^2-1)/(12L)$ for $n=1$, dots and squares are spectral sums for $n=2$ and $n=3$, and the dashed lines are $(\ln L)/2\pi+0.0488$~\cite{Childs2004SpatialWalk} and the three-dimensional limit $0.2527$.
}\label{fig:mechanism}
\end{figure*}}

\newcommand{\mainfiguretwo}{%
\begin{figure*}[!t]
\centering
\includegraphics[width=0.99\textwidth]{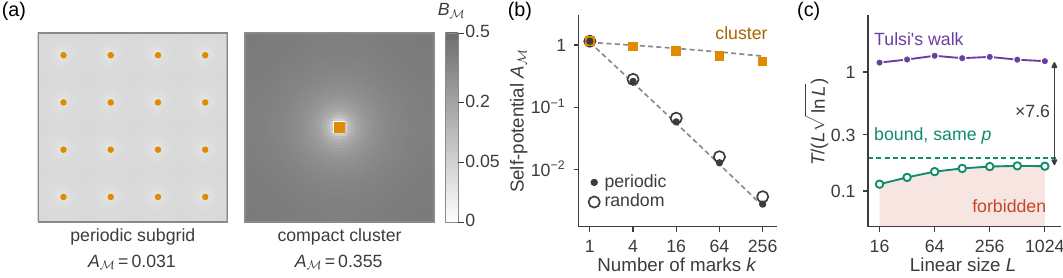}
\caption{%
Marked sets and tightness of the bound.
(a)~Potential drop $B_\Mset(x)=A_\Mset-K_\Mset(x)$ for two arrangements of $k=16$ marked vertices (orange) on a $64\times64$ torus, a periodic $4\times4$ subgrid and a compact $4\times4$ cluster, on a common square-root grey scale. The spatial average of $B_\Mset$ equals $A_\Mset$.
(b)~Self-potential $A_\Mset$ on the $1024\times1024$ torus for periodic subgrids (filled circles), uniformly random sets (open circles, medians of four samples, whose spread is smaller than the symbols) and compact square clusters (orange squares), with $\ln(N/k)/(4\pi k)$ and $\ln(N/k)/(4\pi)$ dashed. At $k=1$ each set is a single marked vertex.
(c)~Number of local steps $T$ at which Tulsi's walk~\cite{Tulsi2008Faster} reaches its peak success probability for one marked vertex from a uniform initial state (filled circles), and the sharpened bound $\Tmin$ at the same success probability (open circles), both divided by $L\sqrt{\ln L}$. The dashed line is the $L\to\infty$ limit $(1-\sqrt{1-p})/\sqrt\pi$ of $\Tmin/(L\sqrt{\ln L})$ at $p=0.558$, the success probability at $L=1024$. No algorithm with the same success probability enters the red region. Lines guide the eye.
}\label{fig:marked}
\end{figure*}}

\makeatletter
\def\frontmatter@thefootnote{\altaffilletter@sw{\@alph}{\@fnsymbol}{\numexpr\csname c@\@mpfn\endcsname+1\relax}}
\makeatother

\begin{document}
\title{Electrostatic Limits on Quantum Spatial Search}

\newcommand{\AffTsinghua}{Center for Quantum Information, IIIS, Tsinghua University, Beijing 100084, China}
\newcommand{\AffHefei}{Hefei National Laboratory, Hefei 230088, China}
\author{Weikang Li}
\email{weikang\_li@tsinghua.edu.cn}
\affiliation{\AffTsinghua}
\affiliation{\AffHefei}

\begin{abstract}
\noindent We prove that a quantum walker cannot search a two-dimensional lattice in Grover's $O(\sqrt N)$ time, resolving the long-standing open question posed by Aaronson and Ambainis in 2003. A walker that hops between neighbouring sites of an $L\times L$ lattice, $N=L^2$, needs $\Omega(\sqrt{N\log N})$ local steps at least to find a marked vertex, even with unlimited memory and free queries. Hence, Tulsi's algorithm is asymptotically optimal. Our proof rests on a surprising correspondence with classical electrostatics. We treat the marked vertex as a point charge and weight the overlap between the walker's states with and without the mark by the potential drop, which vanishes where queries act. Only local steps change this weighted overlap, at a rate limited by the field energy, so the lower bound is set by the self-potential of the charge. In the plane this self-potential grows logarithmically with system size, like the Coulomb potential, leading to the extra root-logarithmic factor. Several marked vertices turn the charge into a conductor, and the lower bound depends on their arrangement through its capacitance.
\end{abstract}
\maketitle

\noindent Grover's algorithm finds one marked item among $N$ with $O(\sqrt N)$ queries, an optimal quadratic speedup assuming that every item is reachable in a single step~\cite{Grover1997Quantum,Bennett1997Strengths,Zalka1999Grovers}. When the items sit on a lattice and a quantum walker can move only between neighbouring sites, information about the marked item must travel through space~\cite{Benioff2002Space}. This spatial search problem is a prototype application of quantum walks~\cite{Shenvi2003Quantum,Childs2004SpatialWalk,Kempe2003Quantum,VenegasAndraca2012Quantum,Portugal2018Quantum} and a natural task for lattice platforms such as photonic waveguide arrays, ultracold atoms and superconducting qubit grids~\cite{Tang2018Experimental,Preiss2015Strongly,Gong2021Quantum}. In three or more dimensions a local walker finds the marked vertex in $O(\sqrt N)$ steps, matching Grover's algorithm~\cite{Ambainis2005Coins,Childs2004SpatialDirac,Aaronson2005Quantum}. On the square lattice the fastest known algorithms take $O(\sqrt{N\log N})$ steps. It has remained open whether a better algorithm can remove the extra factor $\sqrt{\log N}$, or whether it is imposed inevitably by two-dimensional locality~\cite{Aaronson2005Quantum,Childs2004SpatialDirac,Abhijith2019Improving,Burke2026Exact}.

Aaronson and Ambainis introduced the spatial query model and searched the $L\times L$ lattice, $N=L^2$, in $O(\sqrt N(\log N)^{3/2})$ steps~\cite{Aaronson2005Quantum}. 
Subsequent works lowered the cost to $O(\sqrt N\log N)$ by combining amplitude amplification with coined walks, which steer the hops with an internal coin state~\cite{Ambainis2005Coins}, coinless walks~\cite{Portugal2017Quantum,Ambainis2015Spatial}, and continuous-time walks~\cite{Childs2004SpatialDirac,Childs2014Spatial}. Tulsi's walk, which adds one control qubit to the coined walk, reaches $O(\sqrt{N\log N})$ with constant success probability~\cite{Tulsi2008Faster}, and several other walks have since matched this scaling~\cite{Ambainis2013Search,Wong2018Faster,Hoyer2020Analysis,Hoyer2022Spatial}. A second line of work relates the cost of search to classical random walks. Quantum-walk search frameworks~\cite{Szegedy2004Quantum,Magniez2011Search,Belovs2013Electric} led to algorithms that find a marked vertex on any graph in the square root of the classical hitting time, up to logarithmic factors~\cite{Krovi2016Quantum,Ambainis2020Quadratic,Apers2021Unified,Apers2022Quadratic}. On the square lattice this hitting time is of order $N\log N$, which again gives the factor $\sqrt{\log N}$. With $k$ marked vertices, these algorithms take the square root of the hitting time of the marked set, up to logarithmic factors~\cite{Hoyer2017Efficient,Ambainis2020Quadratic}. This hitting time depends on how the marked vertices are arranged, and so does the speed of specific quantum walks~\cite{Nahimovs2018Quantum,Nahimovs2017Adjacent,Roget2023Optimality}. However, no matching lower bound is known for general algorithms. The best one has remained $\Omega(\sqrt N)$ for one marked vertex and for $k=o(N)$~\cite{Bennett1997Strengths,Aaronson2005Quantum}, and stronger bounds cover only specific families of walks~\cite{Childs2004SpatialWalk,Chakraborty2020Optimality,Magniez2012Hitting}.

Resolving the above issue requires a lower bound that holds for every local algorithm. Three challenges stand in the way. First, the walker may carry unbounded memory and apply different position-dependent operations at each step, which rules out analyses based on the spectrum or dispersion relation of a fixed walk operator~\cite{Aharonov2001Quantum,Magniez2012Hitting}. Second, a bound must count local steps rather than queries, because $O(\sqrt N)$ queries suffice on the square lattice if the walker moves enough between them~\cite{Abhijith2019Improving,Li2026Universal}. Third, light-cone and hybrid arguments track how far amplitude can travel~\cite{Aaronson2005Quantum,Lieb1972Finite}, but distance alone gives only $\Omega(\sqrt N)$, the diameter of the lattice.

\mainfigureone

Here, we bridge quantum spatial search and classical electrostatics and prove that every local algorithm needs $\Omega(\sqrt{N\log N})$ local steps to search the square lattice. As a result, Tulsi's algorithm is asymptotically optimal. The key is the overlap between the runs with and without the marked vertex, weighted by the electrostatic potential drop from that vertex (Fig.~\ref{fig:mechanism}). The weight vanishes where queries act, and its field energy limits how much any local step can change the overlap, regardless of the workspace size. On a broad class of vertex-transitive graphs, the bound is of order the square root of the classical hitting time. On these graphs, locality therefore permits at most a quadratic speedup over the simple random walk. With several marked vertices, the charge spreads over a conductor whose capacitance governs the bound. For sparse marked sets on the square lattice, compact clusters are harder to find than periodic subgrids by a factor of order $\sqrt k$ (Fig.~\ref{fig:marked}). The bound exposes a propagation cost that query complexity does not capture. Grover search underlies many quantum algorithms for optimization and machine learning~\cite{Montanaro2016Quantum,Biamonte2017Quantum}, but their implementation cost depends on how data are accessed~\cite{Giovannetti2008Quantum,Aaronson2015Read}. On a two-dimensional lattice, local propagation imposes an unavoidable search cost even when oracle queries are free.

\textit{Model.} Let $G=(V,E)$ be a connected vertex-transitive graph, whose symmetries map any vertex to any other, with $N\ge2$ vertices and degree $d$. A single walker~\cite{Aaronson2005Quantum} has the state space $\HH=\bigoplus_{x\in V}\HH_x$, where a walker at vertex $x$ carries an internal workspace $\HH_x$ of arbitrary dimension. An algorithm starts from a state independent of the marked vertex $m$ and applies two kinds of operations in any order. A local step is a unitary $U$ that may depend on time and position but not on $m$, and whose blocks $U_{xy}\colon\HH_y\to\HH_x$ vanish unless $x=y$ or $\{x,y\}\in E$. A query $O_m$ acts on $\HH_m$ by an arbitrary, possibly time-dependent unitary and as the identity on every other $\HH_x$. A final measurement $\{M_n\}$ outputs a guess $n$, and $p$ denotes the success probability averaged over $m$. With a marked set $\Mset$, queries act in this way on every vertex of $\Mset$, and outputting any of them counts as success. The model contains coined walks, Szegedy's quantized random walks~\cite{Szegedy2004Quantum} and Tulsi's walk directly, and staggered walks at a constant-factor cost. On-site measurements can be deferred into the workspace (Sec.~S1 of the Supplemental Material~\footnote{See Supplemental Material at [URL will be inserted by publisher] for the scope of the model, complete proofs, lattice estimates, the matching algorithms, the sharpened speed limit, continuous-time evolution and numerical illustrations, which includes Refs.~\cite{Aaronson2005Quantum,Bennett1997Strengths,Childs2004SpatialWalk,Krovi2016Quantum,Doyle1984Random,Tetali1991Random,Chandra1996Electrical,Kemeny1976Finite,Levin2017Markov,Lawler2010Random,Tulsi2008Faster,Nahimovs2018Quantum,Roget2023Optimality,Hoyer2017Efficient,Ambainis2020Quadratic,Apers2021Unified,Apers2022Quadratic,Li2026Universal}.}). We count only the number $T$ of local steps and grant queries for free, so the bounds also limit the total time. Finite-size inequalities hold for every $p$, whereas statements with $\Omega$, $\Theta$ or $o(1)$ refer to a fixed target success probability $0<p<1$ (Sec.~S1.3).

\textit{Electrostatic bound.} After $t$ local steps and the queries applied so far, let $a_t$ be the state of the unmarked run, in which every query acts as the identity, and $b_t^m$ the state of the run in which $m$ is marked. For a real weight $w(x,m)$, let $\mathsf W_m$ multiply $\HH_x$ by $w(x,m)$, and define the weighted overlap
\begin{equation}
 F_t=\frac1N\sum_m\bra{a_t}\mathsf W_m\ket{b_t^m}.
 \label{eq:main-overlap}
\end{equation}
It weights the overlap of the hybrid argument~\cite{Bennett1997Strengths} by position, in the spirit of adversary methods~\cite{Ambainis2002Quantum}. It may be complex and need not change monotonically. We require $w(m,m)=0$ and $w(gx,gm)=w(x,m)$ for every automorphism $g$ of $G$, so that the mean $\bar w$ of $w(\cdot,m)$ and its Dirichlet energy $\mathcal E_w=\sum_{\{x,y\}\in E}[w(x,m)-w(y,m)]^2$ do not depend on $m$. Here $\mathcal E_w$ is the field energy when $w$ is viewed as a potential. Since a query changes $b^m$ only inside $\HH_m$, where $\mathsf W_m$ vanishes, $\mathsf W_mO_m=\mathsf W_m$ and queries leave each term of $F$ unchanged. Initially $a_0=b_0^m$, and $F_0=\bar w$ because $\sum_mw(x,m)=N\bar w$ for every $x$ by vertex transitivity.

A local step $U$ changes the $m$th term by $\bra{a}D_m\ket{b^m}$, where $D_m=U^\dagger\mathsf W_mU-\mathsf W_m=U^\dagger[\mathsf W_m,U]$ is Hermitian, so $|\Delta F|\le N^{-1}\sum_m\norm{[\mathsf W_m,U]a}$. This estimate uses the unmarked state alone, so the marked runs may be arbitrary. The commutator $([\mathsf W_m,U]a)_x=\sum_yU_{xy}[w(x,m)-w(y,m)]a_y$ measures how much the weight changes along the edges on which $U$ moves amplitude. Since each block row $(U_{xy})_y$ of a unitary is a contraction, $\norm{([\mathsf W_m,U]a)_x}^2\le\sum_{y\sim x}[w(x,m)-w(y,m)]^2\norm{a_y}^2$. After summation over $x$ and $m$, vertex transitivity makes the coefficient of each $\norm{a_y}^2$ equal to $2\mathcal E_w$, hence
\begin{equation}
 \frac1N\sum_m\norm{[\mathsf W_m,U]a}^2\le\frac{2\mathcal E_w}{N},\qquad |\Delta F|\le\sqrt{\frac{2\mathcal E_w}{N}}
 \label{eq:main-speed}
\end{equation}
for every local step and every normalized $a$ (Fig.~\ref{fig:mechanism}(b)). Averaging over $m$ is essential, because each weight is steep near its own marked vertex but the walker cannot be near every vertex at once. A strong bound needs a large mean and a small energy, two demands that conflict because the weight vanishes at $m$.

\mainfiguretwo

Electrostatics resolves this conflict. Let $\Lap=dI-\mathrm{Adj}$ be the graph Laplacian, with $\mathrm{Adj}$ the adjacency matrix. The potential $K(\cdot,m)$ of a unit charge at $m$ in a uniform neutralizing background is the Green function of the graph,
\begin{equation}
 \Lap K(\cdot,m)=\delta_m-\tfrac1N,\qquad \textstyle\sum_x K(x,m)=0 .
 \label{eq:main-poisson}
\end{equation}
With the Dirichlet form $\langle u,\Lap v\rangle=\sum_{\{x,y\}\in E}[u(x)-u(y)][v(x)-v(y)]$, Eq.~\eqref{eq:main-poisson} gives $\bar w=-\langle w(\cdot,m),\Lap K(\cdot,m)\rangle$ for every weight that vanishes at $m$. It also shows that the energy of the potential equals its value at the charge, the self-potential $A=K(m,m)$,
\begin{equation}
 \sum_{\{x,y\}\in E}[K(x,m)-K(y,m)]^2=A .
 \label{eq:main-energy}
\end{equation}
The Cauchy--Schwarz inequality for the Dirichlet form then implies $\bar w\le\sqrt{\mathcal E_wA}$, with equality only for nonnegative multiples of the potential drop from $m$, $B_m(x)=A-K(x,m)$ (Fig.~\ref{fig:mechanism}(a)). This is Dirichlet's principle. Among all functions that vanish at $m$ and have a given mean, a multiple of $B_m$ has the least energy. With $w=B_m$, $F$ starts at $A$ and moves by at most $\sqrt{2A/N}$ per local step. Physically, $A$ is the energy dissipated when a unit current enters the network at $m$ and leaves uniformly through all vertices~\cite{Doyle1984Random}, and $\thit=NdA$ is the mean hitting time of $m$ for the simple random walk from the uniform distribution, where a walk that starts at $m$ hits it at time zero~\cite{Lovasz1996Random,Levin2017Markov}. Vertex transitivity makes $A$ and the norm $S=[\sum_xK(x,m)^2]^{1/2}$ independent of $m$, and in terms of the nonzero Laplacian eigenvalues $\lambda$, $A=N^{-1}\sum_{\lambda>0}\lambda^{-1}$ and $S^2=N^{-1}\sum_{\lambda>0}\lambda^{-2}$.

It remains to relate $F$ to success. At the end, write $c_m=\braket{a_T}{b_T^m}$ and $\bar c=N^{-1}\sum_mc_m$. With $\KK_m$ multiplying $\HH_x$ by $K(x,m)$, Eq.~\eqref{eq:main-overlap} for $w=B_m$ gives $F_T=A\bar c-r$ with $r=N^{-1}\sum_m\bra{a_T}\KK_m\ket{b_T^m}$, and $|r|\le S/\sqrt N$ because $\sum_mK(x,m)^2=S^2$. Splitting $c_m$ with $M_m$ and $I-M_m$ gives $|c_m|\le\sqrt{q_m}+\sqrt{1-p_m}$, where $p_m=\bra{b_T^m}M_m\ket{b_T^m}$ and the probabilities $q_m=\bra{a_T}M_m\ket{a_T}$ sum to at most one. By concavity of the square root, $\mathrm{Re}\,\bar c\le\sqrt{1-p}+N^{-1/2}$. With Eq.~\eqref{eq:main-speed} for $\mathcal E_w=A$, this proves the electrostatic bound (Sec.~S3).

\begin{theorem*}
On a connected vertex-transitive graph, every algorithm in the spatial query model that succeeds with average probability $p$ uses
\begin{equation}
 T\ \ge\ \frac{\bigl(1-\sqrt{1-p}-N^{-1/2}\bigr)\sqrt N A-S}{\sqrt{2A}}
 \label{eq:main-theorem}
\end{equation}
local steps.
\end{theorem*}

With the endpoint ratio $\eta=S/(\sqrt NA)$, the right side of Eq.~\eqref{eq:main-theorem} is $[1-\sqrt{1-p}-N^{-1/2}-\eta]\sqrt{NA/2}$, so every algorithm needs $\Omega(\sqrt{NA})$ local steps whenever $\eta\to0$. The same bound holds for deciding whether any vertex is marked with error probability at most $\varepsilon<1/2$, with $1-\sqrt{1-p}-N^{-1/2}$ replaced by $1-2\sqrt{\varepsilon(1-\varepsilon)}$, because the unmarked run is then a legitimate input. For continuous-time evolution under nearest-neighbour Hamiltonians $H_t$ with oracle terms on $\HH_m$, only the hopping part moves $F$, and the integrated hopping strength $\int\norm{H_t^{\rm hop}}dt$ obeys the same bound as $T$ (Secs.~S3 and~S8).

\textit{Square lattice.} As in P\'olya's theorem on random walks~\cite{Polya1921}, the plane is the borderline case. The self-potential is $1/2d$ times the expected number of visits of the lazy simple random walk to its starting vertex in excess of the uniform rate $1/N$ per step (Sec.~S2), and on the tori $\mathbb Z_L^n$ it diverges as $L\to\infty$ exactly when the walk on $\mathbb Z^n$ is recurrent, that is, for $n\le2$ (Fig.~\ref{fig:mechanism}(c)). On the $L\times L$ torus, Fourier estimates~\cite{Lawler2010Random} give
\begin{equation}
 A=\frac{\ln L}{2\pi}+O(1),\qquad S\le0.194\,L
 \label{eq:main-torus}
\end{equation}
(Sec.~S4). The growth of $A$ reflects the logarithmic form of the two-dimensional Coulomb potential. Hence $\eta=O(1/\ln L)$, and the theorem becomes
\begin{equation}
 T\ge\bigl(1-\sqrt{1-p}\bigr)\sqrt{\frac{N\ln N}{8\pi}}\,\bigl[1+o(1)\bigr] .
 \label{eq:main-2d}
\end{equation}
Tulsi's walk reaches constant success probability with $O(\sqrt{N\log N})$ local steps~\cite{Tulsi2008Faster}, and a fixed number of verified repetitions reaches any fixed $p<1$, so the optimal number of local steps on the torus is $\Theta(\sqrt{N\log N})$. The same holds on the open lattice. Equation~\eqref{eq:main-2d} still applies because its edges are torus edges, and Tulsi's walk runs on it with at most twice as many local steps, since the torus embeds in it with neighbouring sites at most two steps apart (Sec.~S5). This settles the question of Aaronson and Ambainis~\cite{Aaronson2005Quantum}. The extra factor $\sqrt{\log N}$ is thus set by two-dimensional space, not by the known algorithms, and remains even with only $O(\sqrt N)$ queries~\cite{Abhijith2019Improving}.

The weight $B_m$ has a direct meaning for electrical networks and random walks. Its value $B_m(x)$ is half the effective resistance between $m$ and $x$~\cite{Tetali1991Random,Chandra1996Electrical}, and $NdB_m(x)$ is the mean time that the simple random walk takes to reach $m$ from $x$ (Sec.~S2). Effective resistance already bounds the cost of quantum-walk search algorithms from above~\cite{Belovs2013Electric,Piddock2019Quantum}. Here the effective resistance $A$ between the uniform distribution and $m$ bounds the cost of every local algorithm from below. As $S^2\le A/\lambda_2$, with $\lambda_2$ the smallest nonzero Laplacian eigenvalue, Eq.~\eqref{eq:main-theorem} implies
\begin{equation}
 T\ge\bigl(1-\sqrt{1-p}-N^{-1/2}\bigr)\sqrt{\frac{\thit}{2d}}-\frac1{\sqrt{2\lambda_2}} .
 \label{eq:main-hitting}
\end{equation}
Whenever $\thit\lambda_2/d\to\infty$, every local algorithm thus needs $\Omega(\sqrt{\thit/d})$ local steps. The hitting-time algorithm of Ref.~\cite{Krovi2016Quantum} finds one marked vertex in $O(\sqrt{\thit})$ local steps, including preparation from a known vertex and repetition to any fixed success probability (Sec.~S3), so this scaling is tight for bounded degree. On graphs that are also edge-transitive, with all edges equivalent, the lower bound improves to $\Omega(\sqrt{\thit})$ whenever $\eta\to0$ (Sec.~S7). On tori of dimension $n\ge3$, where $A$ is bounded, it becomes $\Omega(\sqrt N)$. On a cycle, where $\eta\to1/\sqrt5$, a truncated distance weight replaces the potential and gives $\Omega(N)$ for every fixed $p$ (Sec.~S4).

\textit{Several marked vertices.} With $k$ marked vertices the weight must vanish on the whole marked set $\Mset$. Dirichlet's principle then selects the potential of a conductor that occupies $\Mset$, and the point charge spreads into the equilibrium charge of $\Mset$ (Fig.~\ref{fig:marked}(a)). On the $L\times L$ torus, let $K_\Mset$ be the zero-mean potential that takes a constant value $A_\Mset$ on $\Mset$ and solves $\Lap K_\Mset=\rho_\Mset-1/N$ for a unit charge $\rho_\Mset\ge0$ supported on $\Mset$. Its self-potential is the inverse capacitance of the conductor, $A_\Mset=1/[\mathbf 1^{\mathsf T}(\Lap^+)_{\Mset\Mset}^{-1}\mathbf 1]$ with $(\Lap^+)_{\Mset\Mset}$ the block of the Laplacian pseudoinverse on $\Mset$, and it fixes the mean hitting time of $\Mset$ from the uniform distribution, $\thit(\Mset)=4NA_\Mset$. Let the inputs be the $N$ translates of $\Mset$, with $p$ averaged over them. The proof then applies with the translates of $K_\Mset$ in place of $K(\cdot,m)$, which are again Dirichlet-optimal (Sec.~S6). The only change is that $k$ outputs are now correct, which replaces $N^{-1/2}$ by $\sqrt{k/N}$ and gives
\begin{equation}
 T\ge\Bigl[1-\sqrt{1-p}-\sqrt{k/N}-\eta_\Mset\Bigr]\sqrt{\frac{NA_\Mset}{2}} .
 \label{eq:main-marked}
\end{equation}
Here the endpoint ratio $\eta_\Mset=S_\Mset/(\sqrt NA_\Mset)$, with $S_\Mset$ the norm of $K_\Mset$, is the coefficient of variation of the mean hitting time over starting vertices. Equation~\eqref{eq:main-marked} holds for every nonempty proper marked set, even when the algorithm knows the pattern and only its translation is hidden (Sec.~S6).

In the sparse regime $k=o(N)$, Eq.~\eqref{eq:main-marked} and matching algorithms fix the optimal scaling for two opposite arrangements, which differ by a factor of order $\sqrt k$. For a periodic subgrid of spacing $\ell=\sqrt{N/k}$, $K_\Mset$ is $1/k$ times the potential of one marked vertex on the $\ell\times\ell$ quotient torus, where all marked vertices map to a single site. Hence $A_\Mset=\ln(N/k)/(4\pi k)+O(1/k)$ and $\eta_\Mset\to0$. Tulsi's walk matches this bound on the quotient torus, a reduction that Ref.~\cite{Nahimovs2018Quantum} applied to the coined walk, and the optimal number of local steps is $\Theta(\sqrt{(N/k)\log(N/k)})$. A compact cluster instead acts as a single conductor of radius of order $\sqrt k$, whose self-potential in the plane falls only logarithmically with its size: $A_\Mset=\ln(N/k)/(4\pi)+O(1)$, $\eta_\Mset\to0$ and $T\ge(1-\sqrt{1-p})\sqrt{N\ln(N/k)/(8\pi)}\,[1+o(1)]$. Since the worst case over placements is at least this translation average, finding one of $k$ marked vertices takes $\Omega(\sqrt{N\log(N/k)})$ local steps, even when $k$ is known. The same bound holds for deciding between none and exactly $k$ marked vertices with fixed error $0<\varepsilon<1/2$, the problem $\mathrm{OR}^{(k)}$ of Ref.~\cite{Aaronson2005Quantum}, and improves the two-dimensional bound $\Omega(\sqrt N)$ given there. For known $k$, random thinning~\cite{Aaronson2005Quantum} gives a matching upper bound $O(\sqrt{N\log(N/k)})$ for both problems, by keeping one random vertex in each cell of a coarse lattice and searching these vertices with Tulsi's walk (Sec.~S6). On the $1024\times1024$ torus with $k=64$ (Fig.~\ref{fig:marked}(b)), an $8\times8$ cluster has $52$ times the self-potential of the periodic subgrid, whereas four uniformly random sets exceed it by only $24\%$--$26\%$ (Sec.~S9).

On the square torus the speed limit sharpens, which lets us test how much room the bound leaves. Since all its edges are equivalent and it has no triangles for $L\ge4$, each local step moves $F$ by at most $\sqrt{A/2N}$, which doubles the bound~\eqref{eq:main-theorem} to a value $\Tmin$ (Sec.~S7). We simulated Tulsi's walk for $L=16$ to $1024$ (Fig.~\ref{fig:marked}(c)). At $L=1024$ it reaches $p=0.558$ after $T=7.6\,\Tmin$ local steps, and $T/\Tmin$ decreases steadily from $10.5$ at $L=16$. Its overlap $F$, which is real here, moves on average at $0.44$ times the sharpened speed limit but falls $3.4$ times as far as the bound requires, so the gap comes more from the excess fall of $F$ than from its rate (Sec.~S9).

\textit{Discussion.} Ref.~\cite{Magniez2012Hitting} conjectured that reversible quantum walks offer at most a quadratic speedup in hitting time over a suitable classical walk on the same graph. The electrostatic bound establishes an analogous limit for arbitrary local algorithms on a broad class of vertex-transitive graphs, with the simple random walk as the classical reference. Likewise, the gap of order $\sqrt k$ between compact clusters and periodic subgrids in the sparse regime is a property of the search problem rather than of particular walks~\cite{Nahimovs2018Quantum,Roget2023Optimality}.

The hypotheses of the model clarify how additional resources can change the limit. Several walkers or excitations each couple to the oracle~\cite{Lewis2024Quantum}. Long-range hops change the underlying graph and lower its self-potential~\cite{Lewis2021Optimal,Giri2023Quantum,King2025Optimal}. Nonlinear dynamics breaks the linearity and unitarity on which the overlap argument rests~\cite{DiMolfetta2020Searching,Meyer2026Spatial}. Oracle fields that extend beyond the marked vertices can change the weighted overlap directly~\cite{Zylberman2021Dirac}. Interesting questions for future work include matching bounds for an unknown number of marked vertices and for dense or typical random marked sets. Another is the extension to graphs without vertex transitivity, such as fractals and random networks. More broadly, the argument suggests a strategy to prove lower bounds for other local quantum algorithms: choose a weight that vanishes where the oracle acts, minimize its Dirichlet energy at fixed mean, and relate the final overlap to success.

\begin{acknowledgments}
We thank Dong-Ling Deng for his invaluable comments and suggestions. We also thank Da-Wei Wang, Zhide Lu, and Yixuan Ma for discussions.
We acknowledge support from the National Natural Science Foundation of China (T2225008), the Quantum Science and Technology-National Science and Technology Major Project (2021ZD0302203), the Fundamental and Interdisciplinary Disciplines Breakthrough Plan of the Ministry of Education of China (JYB2025XDXM112), and the Tsinghua University Dushi Program.
We acknowledge the use of GPT-6 Astra and Opus-5.5 to assist with theoretical calculations. We checked all derivations and numerical results and take full responsibility for the content.
\end{acknowledgments}

\textit{Data availability.} The data and code supporting the results of this study are available at \href{https://github.com/LWKJJONAK/QuSearch}{\nolinkurl{github.com/LWKJJONAK/QuSearch}}.

\AddToHook{env/thebibliography/begin}{\interlinepenalty=10000}
\bibliography{references}
\end{document}

\title{\texorpdfstring{Supplemental Material:\\Electrostatic Limits on Quantum Spatial Search}{Supplemental Material: Electrostatic Limits on Quantum Spatial Search}}
\hypersetup{pdftitle={Supplemental Material: Electrostatic Limits on Quantum Spatial Search}}

\maketitle
\noindent This Supplemental Material proves the electrostatic bound for quantum spatial search with one marked vertex on vertex-transitive graphs and with translated marked sets on the square torus. It evaluates both bounds on lattices, constructs the matching algorithms and describes the numerical illustrations. Table~\ref{si:notation} collects the notation.

% \tableofcontents

\begin{table}[htbp]
\caption{Notation.}\label{si:notation}
\centering\small
\begin{tabular}{@{}ll@{}}
\toprule
Symbol & Meaning\\ \midrule
$G=(V,E)$, $N$, $d$ & connected vertex-transitive graph, number of vertices ($N\ge2$), degree\\
$\HH=\bigoplus_x\HH_x$ & walker space; $\HH_x$ is the internal workspace at vertex $x$\\
$U$, $U_{xy}$, $U_{\rm d}$ & local step, its blocks $\HH_y\to\HH_x$, its on-site part $\bigoplus_xU_{xx}$\\
$T$ & number of local steps\\
$a_t$, $b_t^m$, $b_t^g$ & unmarked run; marked run with mark $m$; marked run with marked set $\Mset+g$\\
$\{M_n\}$, $p$ & final measurement; success probability averaged over the inputs\\
$\Lap=dI-\mathrm{Adj}$, $\lambda_2$, $\mathcal G$ & graph Laplacian, its smallest nonzero eigenvalue, its pseudoinverse $\Lap^+$\\
$K(x,m)$, $A$, $S$ & potential of a unit charge at $m$ (Eq.~\eqref{si:eq-poisson}), self-potential $K(m,m)$, norm $[\sum_xK(x,m)^2]^{1/2}$\\
$\KK_m$, $\BB_m$ & multiplication by $K(x,m)$ on $\HH_x$; $A-\KK_m$\\
$F_t$ & weighted overlap, Eqs.~\eqref{si:eq-overlap} and~\eqref{si:eq-overlap-marked}\\
$\Mset$, $k$, $g$ & marked set on the $L\times L$ torus, $k=|\Mset|$, translation vector\\
$K_\Mset$, $\rho_\Mset$, $A_\Mset$, $S_\Mset$ & equilibrium potential and charge of $\Mset$, its self-potential and norm\\
$\eta$, $\eta_\Mset$ & endpoint ratios $S/(\sqrt NA)$ and $S_\Mset/(\sqrt NA_\Mset)$\\
$\thit$, $\thit(\Mset)$ & mean hitting time of a vertex, of $\Mset$, from the uniform distribution\\
$\Tmin$ & sharpened bound on the square torus, Eq.~\eqref{si:eq-sharp-bound}\\
\bottomrule
\end{tabular}
\end{table}

\section{Spatial query model}
\label{si:model}
\subsection{Definitions}
Let $G=(V,E)$ be a finite, connected, vertex-transitive simple graph with $N\ge2$ vertices and degree $d$. The state space of a single walker is the direct sum $\HH=\bigoplus_{x\in V}\HH_x$, where the internal workspace $\HH_x$ at vertex $x$ is finite-dimensional and its dimension may grow with $N$. Write $v=(v_x)_{x\in V}$ for a vector of $\HH$, with $v_x\in\HH_x$.

An algorithm starts from a normalized state $a_0$ that does not depend on the input and applies two kinds of operations, interleaved in any order. A \emph{local step} is a unitary $U$ on $\HH$ that does not depend on the input and whose blocks $U_{xy}\colon\HH_y\to\HH_x$ vanish unless $x=y$ or $\{x,y\}\in E$. It may depend on time and position. A \emph{query} with marked vertex $m$ is a unitary $O_m$ that acts on $\HH_m$ by an arbitrary, possibly time-dependent unitary and as the identity on $\HH_x$ for every $x\neq m$. With a marked set $\Mset$, whose vertices are also called marks, the query acts by an arbitrary on-site unitary on every $\HH_x$ with $x\in\Mset$ and as the identity elsewhere. The lower bounds hold for all such queries, and the upper bounds use the usual membership query, which flips an internal flag or a phase at marked vertices. We denote the number of local steps by $T$ and grant queries for free. At the end, a measurement $\{M_n\}_{n\in V}$ that is the same for all inputs, with an optional failure outcome, outputs a vertex $n$, and the algorithm succeeds if $n$ is marked.

The \emph{marked run} $b_t^m$ applies the operations of the algorithm with the queries of input $m$. Section~\ref{si:marked} writes $b_t^g$ for the run on a translated marked set. The \emph{unmarked run} $a_t$ applies the same operations with every query replaced by the identity, and it is defined whether or not the unmarked input belongs to the promised problem. All runs are normalized. Here $t$ counts local steps: $a_t$ and $b_t^m$ are the states after $t$ local steps and the queries applied so far, and $a_T$, $b_T^m$ are the final states. For one marked vertex, the success probability is $p_m=\bra{b_T^m}M_m\ket{b_T^m}$, and $p=N^{-1}\sum_mp_m$.

\subsection{Scope of the model}
Local steps are the $Z$-local transformations of Aaronson and Ambainis~\cite{Aaronson2005Quantum}, in which amplitude moves only along edges. Their $C$-local transformations are special cases. In these, the basis states are partitioned into disjoint subspaces, each supported on one vertex or on a pair of adjacent vertices, and the unitary is block diagonal on this partition, so the operations on different blocks commute. A sequential product of unitaries on overlapping edges can move amplitude across more than one edge and is in general not one local step. The $H$-local transformations $e^{\mathrm iH}$ of Ref.~\cite{Aaronson2005Quantum}, with $H$ nearest-neighbour and $\norm{H}\le\pi$, can also move amplitude across many edges. Each is a unit-time evolution whose hopping part $H-\bigoplus_xH_{xx}$ has norm at most $2\norm{H}\le2\pi$. By Sec.~\ref{si:continuous}, the lower bounds then hold with each such step counted as $2\pi$ local steps. Coined walks (with a coin as workspace), Szegedy walks (with a neighbour label as workspace, and an on-site reflection followed by a move along the labelled edge) and ancilla-controlled walks such as Tulsi's~\cite{Tulsi2008Faster} are local steps interleaved with queries. Staggered walks whose tessellation cells contain vertices at distance two fit the model after each cell unitary is decomposed into a constant number of local steps.

Intermediate on-site measurements can be deferred into the workspace. Their Kraus operators have the form $\Lambda_j=\bigoplus_x\Lambda_j(x)$ with $\sum_j\Lambda_j(x)^\dagger\Lambda_j(x)=I_{\HH_x}$. At each position the isometry $v\mapsto\sum_j\Lambda_j(x)v\otimes\ket{j}$ completes to a unitary on an enlarged workspace, and feedback controlled by the recorded outcomes is a direct sum of local steps and retains block locality. More general intermediate instruments, including global measurements, are admitted only through coherent implementations by local steps, which count towards $T$. Classical random choices independent of the input are purified in the same way.

The bounds concern a fixed number of local steps. For a randomized stopping time $\tau$, measured in operations, suppose its expectation is at most $\mathcal E$ on every input and the average success probability is at least $p_0>0$. Truncation at $\tau_{\max}=\lceil2\mathcal E/p_0\rceil$ operations, with halted branches padded by identities, loses at most $p_0/2$ of the average success probability by Markov's inequality. The bounds applied to the truncated algorithm bound $\tau_{\max}$ from below, and hence $\mathcal E$. No assumption on the stopping time of the unmarked run is needed.

The direct sum over positions describes one walker. The internal workspace may be arbitrarily large and may even store coordinates, but locality and queries refer only to the position $x$. Additional position registers that also couple to the oracle therefore lie outside the model, as do nonlinear evolution and queries acting away from the marked vertices. Long-range hops change the underlying graph rather than leaving the model. The bound then refers to the enlarged graph, whose self-potential must be recomputed, and Theorem~\ref{si:thm-bound} applies when that graph is again vertex-transitive.

\subsection{Range of the results}
\label{si:range}
The finite-size inequalities, Theorems~\ref{si:thm-bound} and~\ref{si:thm-marked} with their corollaries, hold for every success probability $p$. Theorem~\ref{si:thm-marked} holds for every nonempty proper marked set, whether or not the algorithm knows $\Mset$ or $k$. Asymptotic statements need more, since measuring a uniform initial state already succeeds with probability $k/N$ without any local step. Asymptotic lower bounds therefore assume a success probability bounded away from zero, and matching complexities, written with $\Theta$, refer to a fixed target success probability $0<p<1$, or to a fixed error $0<\varepsilon<1/2$ for decision problems. For several marked vertices they refer in addition to the sparse regime $N/k\to\infty$. The forms $\Omega(\sqrt{NA})$ and $\Omega(\sqrt{NA_\Mset})$ hold for every such $p$ when the endpoint ratio $\eta$ or $\eta_\Mset$ tends to zero (Section~\ref{si:bound}), as it does on tori of dimension $n\ge2$, for periodic subgrids and for compact clusters. Section~\ref{si:lattice} treats the cycle, where $\eta$ does not vanish. The matching worst-case upper bound of Theorem~\ref{si:thm-worst} assumes that $k$ is known.

\section{Electrostatic potentials}
\label{si:potential}
\subsection{One charge on a vertex-transitive graph}
The Laplacian $\Lap=dI-\mathrm{Adj}$, with $\mathrm{Adj}$ the adjacency matrix, is positive semidefinite with a simple zero eigenvalue on constant vectors. Write $\Lap^+$ for its Moore--Penrose pseudoinverse and define
\begin{equation}
 K(x,m)=(\Lap^+)_{xm},\qquad\text{so that}\qquad \Lap K(\cdot,m)=\delta_m-\tfrac1N\mathbf 1,\qquad \sum_xK(x,m)=0 .
 \label{si:eq-poisson}
\end{equation}
The function $K(\cdot,m)$ is the electrostatic potential of a unit charge at $m$ in a uniform neutralizing background.

\begin{lemma}[Symmetries]
\label{si:lem-sym}
(i) $K(x,m)=K(m,x)$ and $\sum_mK(x,m)=0$ for every $x$. (ii) $K(gx,gm)=K(x,m)$ for every automorphism $g$ of $G$. (iii) The self-potential $A=K(m,m)$ and $S^2=\sum_xK(x,m)^2$ do not depend on $m$, and
\begin{equation}
 A=\frac1N\sum_{\lambda>0}\frac1\lambda,\qquad S^2=\frac1N\sum_{\lambda>0}\frac1{\lambda^2},
 \label{si:eq-spectral}
\end{equation}
where the sums run over the nonzero eigenvalues of $\Lap$ with multiplicity. In particular $A>0$.
\end{lemma}
\begin{proof}
(i) The pseudoinverse $\Lap^+$ is symmetric and annihilates constants, so its rows and columns sum to zero. (ii) A permutation matrix $P_g$ of an automorphism commutes with $\Lap$, hence with $\Lap^+$. (iii) By (ii) and vertex transitivity the diagonals of $\Lap^+$ and of $(\Lap^+)^2$ are constant, equal to $N^{-1}\operatorname{tr}\Lap^+$ and $N^{-1}\operatorname{tr}(\Lap^+)^2$. Finally, $S^2=\sum_x(\Lap^+)_{xm}^2=[(\Lap^+)^2]_{mm}$.
\end{proof}

\begin{lemma}[Dirichlet energy]
\label{si:lem-energy}
For every $m$,
\begin{equation}
 \sum_{\{x,y\}\in E}[K(x,m)-K(y,m)]^2=A ,
 \label{si:eq-energy}
\end{equation}
and for every vertex $y$,
\begin{equation}
 \sum_m\sum_{x\sim y}[K(x,m)-K(y,m)]^2=2A .
 \label{si:eq-edge-sum}
\end{equation}
\end{lemma}
\begin{proof}
The left side of Eq.~\eqref{si:eq-energy} equals $\langle K(\cdot,m),\Lap K(\cdot,m)\rangle=K(m,m)-N^{-1}\sum_xK(x,m)=A$. Denote the left side of Eq.~\eqref{si:eq-edge-sum} by $\sigma(y)$. For an automorphism $g$, Lemma~\ref{si:lem-sym}(ii) and the substitutions $m=gm'$, $x=gx'$ give $\sigma(gy)=\sigma(y)$, and vertex transitivity makes $\sigma$ constant. Its average is $N^{-1}\sum_m\sum_y\sum_{x\sim y}[K(x,m)-K(y,m)]^2$, in which each edge appears twice, so Eq.~\eqref{si:eq-energy} gives $\sigma=2A$. Edge transitivity is not needed.
\end{proof}

\begin{proposition}[Electrical and random-walk meaning]
\label{si:prop-meaning}
In the network $G$ with unit edge conductances:
(i) $A$ is the Dirichlet energy of the point-to-uniform current flow, in which a unit current enters at $m$ and the currents $1/N$ leave at every vertex. The sink currents are prescribed, and the sink vertices are not held at a common potential. This energy is the effective resistance between $m$ and the uniform distribution.
(ii) The effective resistance between $m$ and $x$ satisfies $R_{\rm eff}(m,x)=2A-2K(m,x)$, and its mean over all vertices $x$, including $m$, is $2A$.
(iii) For the non-lazy simple random walk $P=\mathrm{Adj}/d=I-\Lap/d$, started from the uniform distribution $\pi$, which is stationary, and for $\tau_m=\inf\{t\ge0:X_t=m\}$, the mean hitting times are $\mathbb E_x\tau_m=Nd[A-K(x,m)]$ and $\thit=\mathbb E_\pi\tau_m=NdA$.
(iv) $S^2\le A/\lambda_2$.
(v) For the lazy walk $P'=(I+P)/2$, $A=(2d)^{-1}\sum_{t\ge0}[P'^t(m,m)-N^{-1}]$, that is, $1/2d$ times the expected number of visits of the lazy walk started at $m$ to $m$ in excess of the uniform rate $N^{-1}$ per step.
\end{proposition}
\begin{proof}
(i) The potential $K(\cdot,m)$ drives the net current vector $\delta_m-N^{-1}\mathbf 1$, and its dissipated energy is the left side of Eq.~\eqref{si:eq-energy}~\cite{Doyle1984Random}. (ii) The effective resistance is $R_{\rm eff}(m,x)=(\delta_m-\delta_x)^{\mathsf T}\Lap^+(\delta_m-\delta_x)$~\cite{Tetali1991Random,Chandra1996Electrical}, and averaging over $x$ with Lemma~\ref{si:lem-sym}(i) gives the mean. (iii) The function $h_m(x)=Nd[A-K(x,m)]$ vanishes at $m$ and satisfies $[(I-P)h_m](x)=-N[\Lap K(\cdot,m)](x)=1$ for $x\neq m$. This Dirichlet problem determines $\mathbb E_x\tau_m$ uniquely on a finite connected graph, and averaging over $x$ gives $NdA$. Equivalently, the fundamental matrix $Z=(I-P+\Pi)^{-1}-\Pi$, with $\Pi$ the projector onto constants, equals $d\Lap^+$, and $\mathbb E_\pi\tau_m=Z_{mm}/\pi_m$~\cite{Kemeny1976Finite,Levin2017Markov}, and the inverse exists for periodic chains as well. Counting from $t\ge1$ adds one, conditioning on $X_0\neq m$ divides by $1-1/N$, and a lazy walk doubles the value. (iv) Each term of $S^2$ in Eq.~\eqref{si:eq-spectral} obeys $\lambda^{-2}\le\lambda_2^{-1}\lambda^{-1}$. (v) The lazy chain is aperiodic, so $Z'=\sum_{t\ge0}(P'^t-\Pi)$ converges and equals $(I-P'+\Pi)^{-1}-\Pi$. Since $I-P'=\Lap/2d$, $Z'=2d\,\Lap^+$ and $Z'_{mm}=2dA$.
\end{proof}

\subsection{The equilibrium potential of a marked set}
\label{si:equilibrium}
On the torus $V=\mathbb Z_L^2$, $L\ge3$, $N=L^2$, $\Lap=4I-\mathrm{Adj}$, let $\Mset\subsetneq V$ be nonempty with $k=|\Mset|$, and write $\mathcal G=\Lap^+$ and $\mathcal G_{\Mset\Mset}$ for its principal submatrix indexed by $\Mset$.

\begin{lemma}[Equilibrium potential]
\label{si:lem-equilibrium}
There is a unique zero-mean function $K_\Mset$ that is constant on $\Mset$, with value $A_\Mset$, and satisfies
\begin{equation}
 \Lap K_\Mset=\rho_\Mset-N^{-1}\mathbf 1,\qquad \operatorname{supp}\rho_\Mset\subseteq\Mset,\qquad \sum_x\rho_\Mset(x)=1 .
 \label{si:eq-equilibrium}
\end{equation}
Here $\rho_\Mset(x)\ge N^{-1}$ on $\Mset$, $A_\Mset>0$, and
\begin{equation}
 \sum_{\{x,y\}\in E}[K_\Mset(x)-K_\Mset(y)]^2=A_\Mset,\qquad
 A_\Mset=\frac1{\mathbf 1^{\mathsf T}\mathcal G_{\Mset\Mset}^{-1}\mathbf 1},\qquad
 \rho_\Mset|_\Mset=\frac{\mathcal G_{\Mset\Mset}^{-1}\mathbf 1}{\mathbf 1^{\mathsf T}\mathcal G_{\Mset\Mset}^{-1}\mathbf 1},\qquad K_\Mset=\mathcal G\rho_\Mset .
 \label{si:eq-equilibrium-matrix}
\end{equation}
Equivalently, $A_\Mset=\min\rho^{\mathsf T}\mathcal G\rho$ over charges $\rho$ supported on $\Mset$ with $\mathbf 1^{\mathsf T}\rho=1$, even if signed charges are admitted.
\end{lemma}
\begin{proof}
Solve the Dirichlet problem $u|_\Mset=0$, $(\Lap u)(x)=1$ for $x\notin\Mset$. Since the principal submatrix of $\Lap$ on the unmarked vertices is positive definite, the solution is unique, and the maximum principle ensures $u>0$ off $\Mset$. Set $A_\Mset=N^{-2}\sum_xu(x)$, $K_\Mset=A_\Mset-u/N$ and $\rho_\Mset=N^{-1}\mathbf 1+\Lap K_\Mset$. Then $\sum_xK_\Mset(x)=0$, $K_\Mset|_\Mset=A_\Mset$ and $\rho_\Mset=0$ off $\Mset$. On $\Mset$, $(\Lap u)(s)=-\sum_{y\sim s}u(y)\le0$, so $\rho_\Mset(s)=[1+\sum_{y\sim s}u(y)]/N\ge1/N$, and the total charge is one. The energy is $K_\Mset^{\mathsf T}\Lap K_\Mset=K_\Mset^{\mathsf T}\rho_\Mset=A_\Mset$.

The difference of two solutions is harmonic off $\Mset$, constant on $\Mset$ and of zero mean, so the maximum principle makes it constant everywhere and hence zero. A nonzero vector supported on the proper subset $\Mset$ is not constant, so $\mathcal G_{\Mset\Mset}$ is positive definite. Solving $\mathcal G_{\Mset\Mset}\rho_\Mset=A_\Mset\mathbf 1$ and normalizing gives the matrix formulas, and strict convexity with its Lagrange equations proves the variational statement.
\end{proof}

For $\Mset=\{m\}$ the lemma returns $K(\cdot,m)$ and $A$. Let $\tau_\Mset=\inf\{t\ge0:X_t\in\Mset\}$ for the non-lazy simple random walk $P=\mathrm{Adj}/4$, and $h_\Mset(x)=\mathbb E_x\tau_\Mset$. The Dirichlet equation $(I-P)h_\Mset=1$ off $\Mset$ gives $h_\Mset=4u$, with $u$ from the proof of Lemma~\ref{si:lem-equilibrium}, so
\begin{equation}
 h_\Mset(x)=4N[A_\Mset-K_\Mset(x)],\qquad \thit(\Mset)=\mathbb E_\pi\tau_\Mset=4NA_\Mset .
 \label{si:eq-hit-marked}
\end{equation}
With $S_\Mset^2=\sum_xK_\Mset(x)^2$, the endpoint ratio
\begin{equation}
 \eta_\Mset=\frac{S_\Mset}{\sqrt NA_\Mset}=\frac{\operatorname{sd}_{x\sim\mathrm{Uniform}(V)}h_\Mset(x)}{\mathbb E_{x\sim\mathrm{Uniform}(V)}h_\Mset(x)}
 \label{si:eq-eta}
\end{equation}
is the coefficient of variation of the mean hitting time over starting vertices.

\section{The electrostatic bound}
\label{si:bound}
Let $\KK_m$ multiply $\HH_x$ by $K(x,m)$ and let $\BB_m=A-\KK_m$. The weighted overlap with weight $\BB_m$ is
\begin{equation}
 F_t=\frac1N\sum_m\bra{a_t}\BB_m\ket{b_t^m}.
 \label{si:eq-overlap}
\end{equation}

\begin{lemma}[Speed limit]
\label{si:lem-speed}
For every local step $U$ and every normalized $a\in\HH$,
\begin{equation}
 \frac1N\sum_m\norm{[\KK_m,U]a}^2\le\frac{2A}{N}.
 \label{si:eq-speed}
\end{equation}
\end{lemma}
\begin{proof}
The component of the commutator at vertex $x$ is $([\KK_m,U]a)_x=\sum_yU_{xy}[K(x,m)-K(y,m)]a_y$, where only $y=x$ and the neighbours of $x$ contribute, and the term $y=x$ vanishes. The block row $R_x=(U_{xy})_y$ maps $\bigoplus_y\HH_y$ to $\HH_x$ and satisfies $R_xR_x^\dagger=I_{\HH_x}$, which makes it a contraction. Applied to the vector with components $[K(x,m)-K(y,m)]a_y$, it gives
\begin{equation}
 \norm{([\KK_m,U]a)_x}^2\le\sum_{y\sim x}[K(x,m)-K(y,m)]^2\norm{a_y}^2 .
\end{equation}
Summation over $x$ and $m$ and Eq.~\eqref{si:eq-edge-sum} give $\sum_m\norm{[\KK_m,U]a}^2\le2A\sum_y\norm{a_y}^2=2A$.
\end{proof}

\begin{theorem}[Electrostatic bound]
\label{si:thm-bound}
Every algorithm in the spatial query model on a connected vertex-transitive graph with average success probability $p$ satisfies
\begin{equation}
 T\ \ge\ \frac{\bigl(1-\sqrt{1-p}-N^{-1/2}\bigr)\sqrt N A-S}{\sqrt{2A}} .
 \label{si:eq-bound}
\end{equation}
\end{theorem}
\begin{proof}
\emph{Initial value.} All runs start from $a_0$, and $N^{-1}\sum_m\KK_m=0$ by Lemma~\ref{si:lem-sym}(i), so $F_0=A$.

\emph{Queries.} The query $O_m$ differs from the identity only on $\HH_m$, where $\BB_m$ vanishes since $K(m,m)=A$. Hence $\BB_mO_m=\BB_m$, and a query, which maps $b^m$ to $O_mb^m$ and leaves $a$ unchanged, changes no term of $F$.

\emph{Local steps.} A local step $U$ maps $a\mapsto Ua$ and $b^m\mapsto Ub^m$ and changes the $m$th term by $\bra{a}D_m\ket{b^m}$ with $D_m=U^\dagger\BB_mU-\BB_m=-U^\dagger[\KK_m,U]$. The operator $D_m$ is Hermitian, so $|\bra{a}D_m\ket{b^m}|=|\langle D_ma,b^m\rangle|\le\norm{D_ma}=\norm{[\KK_m,U]a}$. The Cauchy--Schwarz inequality and Lemma~\ref{si:lem-speed} give
\begin{equation}
 |\Delta F|\le\frac1N\sum_m\norm{[\KK_m,U]a}\le\Bigl(\frac1N\sum_m\norm{[\KK_m,U]a}^2\Bigr)^{1/2}\le\sqrt{\frac{2A}N}.
 \label{si:eq-increment}
\end{equation}
After $T$ local steps, therefore, $|F_T-A|\le T\sqrt{2A/N}$.

\emph{Endpoint.} Write $c_m=\braket{a_T}{b_T^m}$ and $\bar c=N^{-1}\sum_mc_m$. Then $F_T=A\bar c-r$ with $r=N^{-1}\sum_m\bra{a_T}\KK_m\ket{b_T^m}$, and
\begin{equation}
 |r|\le\frac1N\sum_m\norm{\KK_ma_T}\le\Bigl(\frac1N\sum_m\sum_xK(x,m)^2\norm{(a_T)_x}^2\Bigr)^{1/2}=\frac S{\sqrt N},
\end{equation}
since $\sum_mK(x,m)^2=S^2$ for every $x$ by Lemma~\ref{si:lem-sym}(i) and (iii).

\emph{Measurement.} Let $q_m=\bra{a_T}M_m\ket{a_T}$, so $\sum_mq_m\le1$. Splitting $c_m=\langle M_m^{1/2}a_T,M_m^{1/2}b_T^m\rangle+\langle(I-M_m)^{1/2}a_T,(I-M_m)^{1/2}b_T^m\rangle$ and applying the Cauchy--Schwarz inequality to each term gives
\begin{equation}
 |c_m|\le\sqrt{p_mq_m}+\sqrt{(1-p_m)(1-q_m)}\le\sqrt{q_m}+\sqrt{1-p_m}.
 \label{si:eq-fidelity}
\end{equation}
The triangle inequality and concavity of the square root give $\mathrm{Re}\,\bar c\le N^{-1}\sum_m(\sqrt{q_m}+\sqrt{1-p_m})\le N^{-1/2}+\sqrt{1-p}$.

\emph{Combination.} Hence $T\sqrt{2A/N}\ge A-\mathrm{Re}\,F_T\ge A(1-\sqrt{1-p}-N^{-1/2})-S/\sqrt N$, which is Eq.~\eqref{si:eq-bound}. A negative right side is a trivial bound.
\end{proof}

\emph{Endpoint ratio.} In terms of the endpoint ratio $\eta=S/(\sqrt NA)$, the right side of Eq.~\eqref{si:eq-bound} is $[1-\sqrt{1-p}-N^{-1/2}-\eta]\sqrt{NA/2}$. For $p\ge p_0>0$ the theorem therefore implies $T=\Omega(\sqrt{NA})$ along every family of graphs with $N\to\infty$ and $\eta\to0$. The ratio need not vanish. On the cycle $\eta\to1/\sqrt5$ (Section~\ref{si:lattice}), and the bound is then of order $\sqrt{NA}$ only for $p>2/\sqrt5-1/5$.

\emph{Query lower bound.} Independently of the graph, consecutive queries merge into at most $Q=T+1$ query blocks $O_{m,j}$, each acting trivially outside $\HH_m$. Let $a_j$ be the unmarked state just before block $j$. The hybrid argument~\cite{Bennett1997Strengths} gives
\[
 \Bigl(\sum_m\norm{b_T^m-a_T}^2\Bigr)^{1/2}
 \le\sum_{j=1}^Q\Bigl(\sum_m\norm{(O_{m,j}-I)a_j}^2\Bigr)^{1/2}\le2Q,
\]
since $\sum_m\norm{(a_j)_m}^2=1$. The final measurement then gives $\sqrt{Np}\le(\sum_m\norm{M_m^{1/2}a_T}^2)^{1/2}+(\sum_m\norm{M_m^{1/2}(b_T^m-a_T)}^2)^{1/2}\le1+2Q$. Thus $T+1\ge(\sqrt{Np}-1)/2$, even with free queries. This bound complements Eq.~\eqref{si:eq-bound} on dense graphs, where the self-potential is small.

\begin{proposition}[Dirichlet-optimal weight]
\label{si:prop-optimal}
Let $f\colon V\times V\to\mathbb R$ satisfy $f(m,m)=0$ and $f(gx,gm)=f(x,m)$ for every automorphism $g$ of $G$, and write $\bar f=N^{-1}\sum_xf(x,m)$ and $\mathcal E_f=\sum_{\{x,y\}\in E}[f(x,m)-f(y,m)]^2$, which do not depend on $m$. With multiplication by $f(\cdot,m)$ in place of $\BB_m$ in Eq.~\eqref{si:eq-overlap}, queries leave the overlap unchanged, its initial value is $\bar f$, and each local step changes it by at most $(2\mathcal E_f/N)^{1/2}$. Moreover $\bar f\le(\mathcal E_fA)^{1/2}$, with equality if and only if $f(\cdot,m)=c\,[A-K(\cdot,m)]$ for some $c\ge0$.
\end{proposition}
\begin{proof}
Invariance and vertex transitivity give $\sum_mf(x,m)=N\bar f$ for every $x$ and $\sum_m\sum_{x\sim y}[f(x,m)-f(y,m)]^2=2\mathcal E_f$ for every $y$, as in Lemma~\ref{si:lem-energy}, and the proofs of Lemma~\ref{si:lem-speed} and Theorem~\ref{si:thm-bound} then apply with $f$ in place of $\BB_m$. For the inequality, $\langle u,\Lap v\rangle=\sum_{\{x,y\}\in E}[u(x)-u(y)][v(x)-v(y)]$. With $u=f(\cdot,m)$ and $v=K(\cdot,m)$, Eq.~\eqref{si:eq-poisson} gives $\langle u,\Lap v\rangle=f(m,m)-\bar f=-\bar f$, and the Cauchy--Schwarz inequality with Lemma~\ref{si:lem-energy} implies $\bar f\le(\mathcal E_fA)^{1/2}$. Equality requires the edge differences of $u$ to be a nonpositive multiple of those of $v$, which on a connected graph means $u=c(A-v)$, since $u(m)=0$.
\end{proof}
Equivalently, $A-K(\cdot,m)$ minimizes the Dirichlet energy among functions that vanish at $m$ and have mean $A$, which is Dirichlet's principle for a grounded point. For a general weight the endpoint term of the bound becomes $\norm{f(\cdot,m)-\bar f}/\sqrt N$, which equals $S/\sqrt N$ for $f=A-K(\cdot,m)$. The proposition thus selects the potential by the ratio $\bar f/\mathcal E_f^{1/2}$, not by the complete finite-size bound, which another weight can improve when the endpoint term is large (Section~\ref{si:lattice}).

\begin{corollary}[Hitting time]
\label{si:cor-hitting}
Under the hypotheses of Theorem~\ref{si:thm-bound},
\begin{equation}
 T\ge\bigl(1-\sqrt{1-p}-N^{-1/2}\bigr)\sqrt{\frac{\thit}{2d}}-\frac1{\sqrt{2\lambda_2}} .
\end{equation}
For fixed $p>0$ it follows that $T=\Omega(\sqrt{\thit/d})$ whenever $\thit\lambda_2/d\to\infty$.
\end{corollary}
\begin{proof}
By Proposition~\ref{si:prop-meaning}(iii), $\sqrt N A/\sqrt{2A}=\sqrt{NA/2}=\sqrt{\thit/2d}$, and by Proposition~\ref{si:prop-meaning}(iv), $S/\sqrt{2A}\le(2\lambda_2)^{-1/2}$.
\end{proof}

\emph{Matching upper bound.} The interpolated walk of Ref.~\cite{Krovi2016Quantum} finds a single marked vertex with constant probability using $O(\sqrt{\thit})$ controlled walk applications. We use the lazy simple random walk. Its hitting time conditioned on an unmarked start, the convention of that reference, is $2\thit/(1-1/N)$. Its stationary state is uniform, and the marked weight $1/N$ and hitting time are independent of the unknown mark. Theorem~6 and Proposition~3 of that reference therefore apply. Each controlled walk is a product of on-site operations, membership queries and a controlled shift along an edge (Sec.~2.4 and Appendix~B.2 of that reference), and costs $O(1)$ local steps here.

Preparation from a known vertex fits within the same budget. Let $D$ be the graph diameter. The variational formula for effective resistance, applied to $f(z)=\operatorname{dist}(x,z)$, gives $\operatorname{dist}(x,y)^2\le |E|R_{\rm eff}(x,y)$, since $\sum_{\{u,v\}\in E}[f(u)-f(v)]^2\le|E|$. Positive semidefiniteness of $\Lap^+$ implies $|K(x,y)|\le A$, and Proposition~\ref{si:prop-meaning} then yields
\[
 D^2\le4|E|A=2\thit.
\]
A breadth-first spanning tree prepares the uniform state in $O(D)$ local steps. At each depth, a vertex with subtree size $n_v$ retains a fraction $1/n_v$ of its incoming probability and sends fractions $n_c/n_v$ to its children $c$, using separate internal labels for retained amplitude and for each outgoing tree edge. All moves at one depth occur in parallel. After a failed trial, returning to the root and preparing again also costs $O(D)$. Verification and a fixed number of repetitions reach any fixed success probability below one, with total cost $O(\sqrt{\thit})$.

\begin{proposition}[Detection]
\label{si:prop-detection}
Suppose an algorithm decides whether a marked vertex exists, accepting the unmarked input with probability at most $\varepsilon$ and every input with one marked vertex with probability at least $1-\varepsilon$, where $0\le\varepsilon<1/2$. Then
\begin{equation}
 T\ge\frac{\bigl(1-2\sqrt{\varepsilon(1-\varepsilon)}\bigr)\sqrt N A-S}{\sqrt{2A}} .
\end{equation}
\end{proposition}
\begin{proof}
The proof of Theorem~\ref{si:thm-bound} applies up to the measurement step. Let $\{M_{\rm yes},M_{\rm no}\}$ be the final measurement, $q=\bra{a_T}M_{\rm yes}\ket{a_T}\le\varepsilon$ and $p_m=\bra{b_T^m}M_{\rm yes}\ket{b_T^m}\ge1-\varepsilon$. As in Eq.~\eqref{si:eq-fidelity}, $|c_m|\le\sqrt{qp_m}+\sqrt{(1-q)(1-p_m)}$. With $\sin^2\theta_q=q$ and $\sin^2\theta_m=p_m$ this equals $\cos(\theta_m-\theta_q)$, and on the region $q\le\varepsilon$, $p_m\ge1-\varepsilon$ the angle difference is at least $\pi/2-2\arcsin\sqrt\varepsilon$, so $|c_m|\le2\sqrt{\varepsilon(1-\varepsilon)}$. Hence $\mathrm{Re}\,\bar c\le2\sqrt{\varepsilon(1-\varepsilon)}$, and the combination step completes the proof.
\end{proof}

\section{Lattices}
\label{si:lattice}
On the torus $\mathbb Z_L^n$, $L\ge3$, with $N=L^n$ and degree $d=2n$, the Laplacian eigenvalues are $\lambda_\nu=\sum_{r=1}^n4\sin^2(\pi\nu_r/L)$ for $\nu\in\mathbb Z_L^n$, and
\begin{equation}
 K(x,0)=\frac1N\sum_{\nu\neq0}\frac{e^{2\pi\mathrm i\nu\cdot x/L}}{\lambda_\nu},\qquad A=\frac1N\sum_{\nu\neq0}\frac1{\lambda_\nu},\qquad S^2=\frac1N\sum_{\nu\neq0}\frac1{\lambda_\nu^2}.
\end{equation}

\begin{proposition}[Square torus]
\label{si:prop-square}
For $n=2$ and $L\ge3$, with $j_0=\lfloor(L-1)/2\rfloor$,
\begin{equation}
 \frac1{\pi^2}\sum_{j=1}^{j_0}\frac1j\ \le\ A\ \le\ \frac12\sum_{j=1}^{\lfloor L/2\rfloor}\frac1j,\qquad S^2\le\frac{\zeta(3)}{32}L^2,
 \label{si:eq-square-bounds}
\end{equation}
and $A=(2\pi)^{-1}\ln L+O(1)$.
\end{proposition}
\begin{proof}
Represent each frequency once by $\nu=(\nu_1,\nu_2)\in\{-\lfloor L/2\rfloor,\dots,\lceil L/2\rceil-1\}^2$, with $r^2=|\nu|^2$. The inequalities $2\theta/\pi\le\sin\theta\le\theta$ on $[0,\pi/2]$ give $16r^2/L^2\le\lambda_\nu\le4\pi^2r^2/L^2$. The Chebyshev shell $\max(|\nu_1|,|\nu_2|)=j$ has $8j$ points when complete, which holds for $j\le j_0$, and at most $8j$ points otherwise, and on it $j^2\le r^2\le2j^2$. The lower bound on $A$ uses the complete shells with $\lambda_\nu\le8\pi^2j^2/L^2$, the upper bound uses all shells with $\lambda_\nu\ge16j^2/L^2$, and $S^2\le N^{-1}\sum_j8j\,L^4/(256j^4)\le\zeta(3)L^2/32$.

For the asymptotics, $0\le u-\sin u\le u^3/6$ and $u^2-\sin^2u\le u^4/3$ on $[0,\pi/2]$ give $0\le4\pi^2r^2/L^2-\lambda_\nu\le4\pi^4r^4/(3L^4)$, and with $\lambda_\nu\ge16r^2/L^2$ this yields the uniform estimate
\begin{equation}
 0\le\frac1{L^2\lambda_\nu}-\frac1{4\pi^2r^2}\le\frac{\pi^2}{48L^2}
 \label{si:eq-inverse-symbol}
\end{equation}
for all $\nu\neq0$. Summation over at most $L^2$ frequencies gives $A=(4\pi^2)^{-1}\sum_{\nu\neq0}|\nu|^{-2}+E_L$ with $0\le E_L\le\pi^2/48$. Comparing $|\nu|^{-2}$ with its integral over the unit cell centred at $\nu$ has error $O(|\nu|^{-3})$, whose lattice sum converges, and the integral of $|x|^{-2}$ over the frequency square with a fixed neighbourhood of the origin removed is $2\pi\ln L+O(1)$~\cite{Lawler2010Random}. Hence $A=(2\pi)^{-1}\ln L+O(1)$.
\end{proof}

For $p\ge p_0>0$, set $\alpha_0=1-\sqrt{1-p_0}$. The right side of Eq.~\eqref{si:eq-bound} is then at least $\alpha_0L\sqrt{A/2}-\sqrt{A/2}-S/\sqrt{2A}$, whose positive term is $\Theta(L\sqrt{\log L})$ and whose subtracted terms are $O(\sqrt{\log L})$ and $O(L/\sqrt{\log L})$. With $A=(2\pi)^{-1}\ln L\,[1+o(1)]$ and $L^2\ln L=\tfrac12N\ln N$, this gives main-text Eq.~(7),
\begin{equation}
 T\ge\bigl(1-\sqrt{1-p}\bigr)\sqrt{\frac{N\ln N}{8\pi}}\,[1+o(1)] .
\end{equation}
On the square torus, $A$ is the sum $S_{1,2}$ of Ref.~\cite{Childs2004SpatialWalk}, whose Eq.~(42) gives $A=(2\pi)^{-1}\ln L+c_\square+O(L^{-2})$ with $c_\square\approx0.0488$. The spectral sums of Eq.~\eqref{si:eq-spectral} up to $L=8192$ are consistent with this expansion and give $c_\square\approx0.048766$ (Fig.~1(c) of the main text). Those up to $L=4096$ give $S/L\approx0.0622$ (Table~\ref{si:tab-constants}).

\begin{table}[htbp]
\caption{Self-potential $A$ and norm $S$ on the torus $\mathbb Z_L^n$, evaluated from the spectral sums of Eq.~\eqref{si:eq-spectral}. For $n=1$, $A=(L^2-1)/(12L)$ exactly.}\label{si:tab-constants}
\centering\small
\begin{tabular}{@{}rrrr@{\qquad}rrrr@{\qquad}rrr@{}}
\toprule
\multicolumn{3}{c}{$n=1$} && \multicolumn{3}{c}{$n=2$} && \multicolumn{3}{c}{$n=3$}\\
$L$ & $A$ & $S$ && $L$ & $A$ & $S$ && $L$ & $A$ & $S$\\ \midrule
16 & 1.3281 & 2.431 && 16 & 0.48992 & 1.0242 && 8 & 0.22461 & 0.3085\\
64 & 5.3320 & 19.104 && 64 & 0.71067 & 3.9907 && 16 & 0.23863 & 0.4253\\
256 & 21.333 & 152.66 && 256 & 0.93131 & 15.923 && 32 & 0.24568 & 0.5926\\
1024 & 85.333 & 1221.2 && 1024 & 1.15194 & 63.678 && 64 & 0.24920 & 0.8311\\
4096 & 341.33 & 9769.5 && 4096 & 1.37258 & 254.71 && 128 & 0.25097 & 1.1704\\
\bottomrule
\end{tabular}
\end{table}

\emph{Other dimensions.} On the cycle ($n=1$), $A=(N^2-1)/(12N)$ and $S^2=N^3/720\,[1+o(1)]$, and Eq.~\eqref{si:eq-bound} becomes $T\ge N\bigl[\tfrac{\sqrt6}{12}(1-\sqrt{1-p})-\tfrac1{\sqrt{120}}\bigr]-O(\sqrt N)$, with $\sqrt6/12\approx0.2041$ and $1/\sqrt{120}\approx0.0913$. The bound is linear in $N$ for $p>2/\sqrt5-1/5\approx0.6944$. Another weight removes this threshold. With $\alpha=1-\sqrt{1-p}$ and $f(x,m)=\min\{\mathrm{dist}(x,m)/r,1\}$ for an integer $r$ with $1\le r<N/2$, direct summation gives $\bar f=1-r/N$, $\mathcal E_f=2/r$ and $\norm{f(\cdot,m)-\bar f}^2=(2r^2+1)/(3r)-r^2/N$. Proposition~\ref{si:prop-optimal} and the remark after it then give $T\ge\tfrac12\sqrt{rN}\,\{(\alpha-N^{-1/2})(1-r/N)-[(2r^2+1)/(3rN)-r^2/N^2]^{1/2}\}$. For $r=\lceil\alpha^2N/16\rceil$ this yields $\liminf_{N\to\infty}T/N\ge(\alpha^2/8)[1-\alpha^2/16-(1/24-\alpha^2/256)^{1/2}]\ge11\alpha^2/128$, since $\alpha\le1$. Hence $T=\Omega(N)$ for every fixed $p>0$.

For $n=3$, $A$ converges to the finite constant
\begin{equation}
 A_\infty:=\frac1{(2\pi)^3}\int_{[-\pi,\pi]^3}\frac{d^3q}{6-2\sum_{i=1}^3\cos q_i}\approx0.2527 .
\end{equation}
The singularity $O(|q|^{-2})$ of the integrand at the origin is integrable. By shell counting, the parts of the spectral sum and of the integral inside a ball of radius $\epsilon$ are both $O(\epsilon)$, uniformly in $L$, and outside the ball the Riemann sum converges. Also $S^2=O(L)$, and for fixed $p>0$ the bound is $T\ge(1-\sqrt{1-p})\sqrt{NA_\infty/2}\,[1+o(1)]$. For every fixed $n\ge3$, $A$ converges in the same way to a positive constant, and shell counting gives $S^2\le C_nL^{4-n}\sum_{1\le j\le L/2}j^{n-5}$, which is $O(L)$, $O(\log L)$ and $O(1)$ for $n=3$, $n=4$ and $n\ge5$. Hence $S=o(\sqrt N)$, and the bound is $\Omega(\sqrt N)$. For $n=3$, $A_\infty=G_3/6$, where $G_3\approx1.5164$ is Watson's integral, the expected number of visits of the simple random walk on $\mathbb Z^3$ to its starting point. In one and two dimensions this walk is recurrent and $A$ diverges, as Proposition~\ref{si:prop-meaning}(v) requires. Two dimensions are the marginal case, in which $A$ diverges, but only logarithmically.

\section{Routing and open boundaries}
\label{si:routing}
\begin{lemma}[Routing]
\label{si:lem-routing}
Let $W$ be a permutation of a set of distinguished basis states of $\HH$, and suppose each distinguished state $e$ comes with a fixed nearest-neighbour path of length at most $D\ge1$ from its vertex to the vertex of $W(e)$, chosen independently of the input. With additional on-site workspace, $W$ can be implemented on the distinguished states by $D$ local steps that leave no amplitude in the added workspace. An on-site unitary $C$ applied before $W$ can be absorbed into the first of these steps.
\end{lemma}
\begin{proof}
Pad each path with on-site waits to exactly $D$ moves, and introduce distinct transit states $\tau_{e,1},\dots,\tau_{e,D-1}$ at its successive intermediate vertices. Define a permutation $R$ of the enlarged basis by $R(e)=\tau_{e,1}$, $R(\tau_{e,j})=\tau_{e,j+1}$ and $R(\tau_{e,D-1})=W(e)$, fixing every other basis state. For $D=1$, put $R=W$. Each distinguished target has exactly one preimage since $W$ is a permutation, and each transit state has exactly one preimage, so $R$ is unitary. Every move crosses at most one edge, so $R$ is a local step. On distinguished states $R^D=W$, and no transit amplitude remains after the $D$ moves. For an on-site $C$, the first step $R(C\oplus I)$ is still local, so $WC$ costs $D$ local steps. Finally, $R$ does not depend on the input, because paths and transit labels are fixed in advance.
\end{proof}

The open $L\times L$ lattice has a subset of the torus edges and the same vertex set. A local step on the open lattice is therefore a local step on the torus, and every open-lattice algorithm obeys Eq.~\eqref{si:eq-bound} with the torus constants. The open lattice is not vertex-transitive, but the proof uses only the symmetry of the torus potential.

\begin{proposition}[Folded torus]
\label{si:prop-fold}
For $L\ge3$, Tulsi's walk on the $L\times L$ torus runs on the open $L\times L$ lattice with at most twice as many local steps, the same queries and at most four additional internal states per vertex.
\end{proposition}
\begin{proof}
With $h=\lceil L/2\rceil$, place cycle vertex $c\in\mathbb Z_L$ at position $\phi(c)=2c$ of the path $\{0,\dots,L-1\}$ for $c<h$ and at $\phi(c)=2(L-c)-1$ for $c\ge h$. The cycle then visits all even positions in increasing order and the odd positions in decreasing order, so every cycle edge joins positions at distance one or two. The product map $\Phi=\phi\times\phi$ is a bijection onto the open lattice that sends every torus edge to a fixed path of length one or two along one axis.

Tulsi's local step is $WC$, an on-site unitary $C$ (the coin and the control) followed by the flip-flop shift $W$, which moves each basis state of the control branch $\ket1$ along one torus edge and fixes the branch $\ket0$. To obtain the stated workspace bound, use two local permutations $P_1,P_2$ without padding the short routes. For a folded edge of length two, let $e,f$ be its flip-flop basis states, with $W(e)=f$ and $W(f)=e$, and add distinct states $\tau_e,\tau_f$ at its midpoint. In cycle notation, set
\[
 P_1=(e\,\tau_e)(f\,\tau_f),\qquad P_2=(e\,\tau_e\,f\,\tau_f).
\]
Then $P_2P_1$ swaps $e$ and $f$ and fixes both transit states, and every move of either permutation crosses at most one open-lattice edge. On each length-one edge, take $P_1=I$ and let $P_2$ perform the flip-flop swap. Both permutations fix the control branch $\ket0$. Because the blocks have disjoint basis labels even when their physical paths intersect, their direct sums are local unitaries and $P_2P_1=W\oplus I$. Absorbing $C\oplus I$ into the first step gives $P_2[P_1(C\oplus I)]$.

For both odd and even $L$, the length-two edges of the folded cycle are exactly $\{j,j+2\}$, $0\le j\le L-3$. Thus a lattice vertex is the midpoint of at most one such edge per axis, which requires at most four added internal states in total. Queries act at $\Phi(m)$ between completed simulated steps, when the transit states are empty, and the final outcome is relabelled by $\Phi^{-1}$.
\end{proof}
An initial state concentrated on one vertex reaches the uniform superposition of Tulsi's walk with $2(L-1)$ further local steps, by preparing a uniform row with neighbouring two-level rotations and then each column in parallel, which does not change the scaling. A membership query verifies the output, and after a failed trial, returning to the starting vertex and preparing again costs $O(L)$ local steps. Since each trial succeeds with probability bounded below by a constant, a fixed number of independent trials reaches any fixed success probability below one, on the torus and on the open lattice, within $O(\sqrt{N\log N})$ local steps.

\section{Several marked vertices}
\label{si:marked}
On the $L\times L$ torus, an input is a translate $\Mset_g=\Mset+g$ of a fixed nonempty proper marked set $\Mset$, with $g\in V$ uniformly distributed. The algorithm may know $\Mset$, and hence $k$, but not $g$. Distinct translations may give the same set, and averaging over all $N$ translations is then equivalent to averaging over the distinct sets of the orbit. Success means outputting any vertex of $\Mset_g$.

\subsection{The bound for translated marked sets}
With $K_\Mset$, $A_\Mset$ and $S_\Mset$ from Section~\ref{si:equilibrium}, let $K_g(x)=K_\Mset(x-g)$, let $\KK_g$ multiply $\HH_x$ by $K_g(x)$, put $\BB_g=A_\Mset-\KK_g$, and define
\begin{equation}
 F_t=\frac1N\sum_g\bra{a_t}\BB_g\ket{b_t^g},
 \label{si:eq-overlap-marked}
\end{equation}
where $b_t^g$ is the run on input $\Mset_g$.
\begin{theorem}[Translated marked sets]
\label{si:thm-marked}
Every algorithm with average success probability $p$ over the translates of $\Mset$ obeys
\begin{equation}
 T\ge\frac{\bigl(1-\sqrt{1-p}-\sqrt{k/N}\bigr)\sqrt NA_\Mset-S_\Mset}{\sqrt{2A_\Mset}}
 =\Bigl(1-\sqrt{1-p}-\sqrt{k/N}-\eta_\Mset\Bigr)\sqrt{\frac{NA_\Mset}{2}} .
 \label{si:eq-marked-bound}
\end{equation}
If $p\ge p_0>0$, $k/N\to0$ and $\eta_\Mset\to0$ along a family of marked sets, then $T=\Omega(\sqrt{NA_\Mset})=\Omega(\sqrt{\thit(\Mset)})$.
\end{theorem}
\begin{proof}
The translation average of $K_g(x)$ is zero for each $x$, so $F_0=A_\Mset$. Since $\BB_g$ vanishes on every vertex of $\Mset_g$, $\BB_gO_g=\BB_g$ for the query $O_g$ of input $\Mset_g$, and queries leave $F$ invariant. The quantity $\sigma(y)=\sum_g\sum_{x\sim y}[K_g(x)-K_g(y)]^2$ is translation invariant, hence constant in $y$, and its average over $y$ counts each edge twice, so the energy identity of Lemma~\ref{si:lem-equilibrium} gives $\sigma(y)=2A_\Mset$. The block-row argument of Lemma~\ref{si:lem-speed} then bounds $N^{-1}\sum_g\norm{[\KK_g,U]a}^2$ by $2A_\Mset/N$, and as in Theorem~\ref{si:thm-bound}, $|F_T-A_\Mset|\le T\sqrt{2A_\Mset/N}$.

At the end, $c_g=\braket{a_T}{b_T^g}$ and $F_T=A_\Mset\bar c-r$, with $r=N^{-1}\sum_g\bra{a_T}\KK_g\ket{b_T^g}$. Translation averaging gives $\sum_gK_g(x)^2=S_\Mset^2$ and hence $|r|\le S_\Mset/\sqrt N$. Let $E_g=\sum_{x\in\Mset_g}M_x$, $q_g=\bra{a_T}E_g\ket{a_T}$ and $p_g=\bra{b_T^g}E_g\ket{b_T^g}$. Splitting the overlap with $E_g$ and $I-E_g$ yields $|c_g|\le\sqrt{q_g}+\sqrt{1-p_g}$. Each vertex belongs to exactly $k$ of the translated sets, counted with multiplicity, so $N^{-1}\sum_gq_g=(k/N)\sum_x\bra{a_T}M_x\ket{a_T}\le k/N$. Concavity gives $\mathrm{Re}\,\bar c\le\sqrt{k/N}+\sqrt{1-p}$, and the final step of the proof of Theorem~\ref{si:thm-bound} proves Eq.~\eqref{si:eq-marked-bound}. Its hitting-time form follows from Eq.~\eqref{si:eq-hit-marked}.
\end{proof}
The translates of $A_\Mset-K_\Mset$ are again the Dirichlet-optimal weights. For a weight $f(x-g)$ that vanishes on $\Mset_g$, Eq.~\eqref{si:eq-equilibrium} gives $\langle f(\cdot-g),\Lap K_g\rangle=\langle f(\cdot-g),\rho_{\Mset_g}\rangle-\bar f=-\bar f$, and $\bar f\le(\mathcal E_fA_\Mset)^{1/2}$ follows as in Proposition~\ref{si:prop-optimal}, with equality only for nonnegative multiples of $A_\Mset-K_g$.

\begin{corollary}[Detection]
\label{si:cor-marked-detection}
Deciding between no marked vertex and a translate of $\Mset$, with false-positive and false-negative probabilities at most $\varepsilon$ on every input, $0\le\varepsilon<1/2$, requires $T\ge[1-2\sqrt{\varepsilon(1-\varepsilon)}-\eta_\Mset]\sqrt{NA_\Mset/2}$.
\end{corollary}
\begin{proof}
For the common yes/no measurement, $q=\bra{a_T}M_{\rm yes}\ket{a_T}\le\varepsilon$ as in Proposition~\ref{si:prop-detection} and $p_g\ge1-\varepsilon$, so the binary fidelity bound of that proposition gives $|c_g|\le2\sqrt{\varepsilon(1-\varepsilon)}$. The remaining steps are unchanged.
\end{proof}

For a random pattern, the theorem covers one realization that the algorithm knows, with its translates as inputs. This differs from drawing a new random set for each input. An asymptotic statement for random patterns would require probabilistic control of both $A_\Mset$ and $\eta_\Mset$.

\subsection{Periodic subgrids}
Let $\ell\ge3$ divide $L$, and let $\Mset=\{(i\ell,j\ell):0\le i,j<L/\ell\}$, with $k=(L/\ell)^2$ and $N/k=\ell^2$. Write $K^{(\ell)}$, $A^{(\ell)}$ and $S^{(\ell)}$ for the single-mark potential, self-potential and norm on the $\ell\times\ell$ torus, the quotient torus of $\Mset$.
\begin{proposition}[Periodic subgrids]
\label{si:prop-periodic}
The equilibrium charge and potential are $\rho_\Mset=k^{-1}\mathbf 1_\Mset$ and $K_\Mset(x)=k^{-1}K^{(\ell)}(x\bmod\ell)$, with $A_\Mset=A^{(\ell)}/k$ and $S_\Mset^2=(S^{(\ell)})^2/k$. Consequently, Eq.~\eqref{si:eq-marked-bound} coincides with the single-mark bound on the quotient torus,
\begin{equation}
 T\ge\frac{\bigl(1-\sqrt{1-p}-\ell^{-1}\bigr)\ell A^{(\ell)}-S^{(\ell)}}{\sqrt{2A^{(\ell)}}}.
\end{equation}
As $\ell\to\infty$, equivalently $N/k\to\infty$,
\begin{equation}
 A_\Mset=\frac{\ln(N/k)}{4\pi k}+O(k^{-1}),\qquad \eta_\Mset=O\bigl(1/\ln(N/k)\bigr),
\end{equation}
and the optimal number of local steps is $\Theta(\sqrt{(N/k)\ln(N/k)})$ for any fixed target success probability $0<p<1$.
\end{proposition}
\begin{proof}
Applying $\Lap$ to the repeated quotient function gives the source $1/k-1/(k\ell^2)=1/k-1/N$ at each marked vertex and $-1/N$ elsewhere. The repeated function has zero mean and is constant on $\Mset$, so uniqueness in Lemma~\ref{si:lem-equilibrium} proves the formula. The norm follows because there are $k$ copies of the quotient. With $A^{(\ell)}=(\ln\ell)/(2\pi)+O(1)$ and $S^{(\ell)}=O(\ell)$ from Proposition~\ref{si:prop-square}, and $\ln\ell=\tfrac12\ln(N/k)$, the asymptotics follow, and substitution in Eq.~\eqref{si:eq-marked-bound} cancels the factors $k^{-1/2}$. For the upper bound, a uniform initial state and identical on-site operations keep the walk in the subspace invariant under translations by $\ell$, which is the quotient torus, and the multi-mark query acts there as a single-mark query. Tulsi's walk~\cite{Tulsi2008Faster} therefore succeeds with constant probability after $O(\ell\sqrt{\ln\ell})$ local steps. Reference~\cite{Nahimovs2018Quantum} uses the same symmetry reduction for the walk of Ambainis, Kempe and Rivosh. A localized start is prepared with $O(\ell)$ local steps in an open $\ell\times\ell$ window containing one marked vertex, where Proposition~\ref{si:prop-fold} runs the walk. A membership query verifies the output, and returning and preparing again within the same window costs $O(\ell)$ local steps per trial, so a fixed number of independent trials reaches any fixed success probability below one.
\end{proof}

For this pattern, and for every pattern that is invariant under exchanging the two axes up to a translation, the correlation $\sum_zK_\Mset(z)K_\Mset(z+e)$ takes the same value for the four unit vectors $e$. The proof of Proposition~\ref{si:prop-identity} then applies to the translates and gives, for $L\ge4$, $N^{-1}\sum_g\norm{[\KK_g,U]a}^2=(A_\Mset/2N)(1-\norm{U_{\rm d}a}^2)\le A_\Mset/2N$, which doubles the right side of Eq.~\eqref{si:eq-marked-bound}. For periodic subgrids the leading term becomes $(1-\sqrt{1-p})\sqrt{(N/k)\ln(N/k)/(2\pi)}$, with a correction of order $\sqrt{(N/k)/\ln(N/k)}$.

\subsection{Compact clusters and the worst case}
For any marked set, positivity of $\rho_\Mset$ and the triangle inequality give
\begin{equation}
 S_\Mset=\Bigl\lVert\sum_{s\in\Mset}\rho_\Mset(s)\mathcal G(\cdot,s)\Bigr\rVert\le S=O(L),
 \label{si:eq-norm-all}
\end{equation}
where $S=\norm{\mathcal G(\cdot,0)}$ is the single-mark norm of Proposition~\ref{si:prop-square}.

\begin{proposition}[Compact clusters]
\label{si:prop-compact}
Let $k\ge1$ and $\kappa=\lfloor\sqrt k\rfloor$, and let the cluster $\mathcal C_k$ have exactly $k$ vertices, contain a $\kappa\times\kappa$ square and lie in a $(\kappa+1)\times(\kappa+1)$ square. Then, uniformly as $N/k\to\infty$,
\begin{equation}
 A_{\mathcal C_k}=\frac{\ln(N/k)}{4\pi}+O(1),\qquad \eta_{\mathcal C_k}=O\bigl(1/\ln(N/k)\bigr),
\end{equation}
and every local algorithm that finds a marked vertex with fixed average success probability $p>0$ on the translates of $\mathcal C_k$ obeys
\begin{equation}
 T\ge\bigl(1-\sqrt{1-p}\bigr)\sqrt{\frac{N\ln(N/k)}{8\pi}}\,[1+o(1)] .
 \label{si:eq-compact-bound}
\end{equation}
\end{proposition}
\begin{proof}
For a unit charge $\rho$, $\rho^{\mathsf T}\mathcal G\rho=L^{-2}\sum_{\nu\neq0}|\widehat\rho(\nu)|^2/\lambda_\nu$, with the frequencies of Proposition~\ref{si:prop-square}. For the lower bound, the equilibrium charge is a probability distribution on a set of diameter $O(\kappa)$. After its support is translated to the origin, $\cos t\ge1-t^2/2$ bounds the deficit $1-|\widehat\rho_{\mathcal C_k}(\nu)|^2$ by $C_0\kappa^2|\nu|^2/L^2$, for an absolute constant $C_0$. Restricting the sum to $0<|\nu|\le c_0L/\kappa$ for a fixed small $c_0>0$ and using $1/(L^2\lambda_\nu)\ge1/(4\pi^2|\nu|^2)$ yields $A_{\mathcal C_k}\ge(4\pi^2)^{-1}\sum_{0<|\nu|\le c_0L/\kappa}|\nu|^{-2}-O(1)=(2\pi)^{-1}\ln(L/\kappa)-O(1)$, since the deficit contributes $O(\kappa^2/L^2)$ for each of $O((L/\kappa)^2)$ frequencies.

For the upper bound, the variational principle of Lemma~\ref{si:lem-equilibrium} with the uniform charge on the contained $\kappa\times\kappa$ square gives a Fourier transform $\varphi_\kappa(\nu_1)\varphi_\kappa(\nu_2)$ with $|\varphi_\kappa(j)|\le\min\{1,L/(2\kappa|j|)\}$ for $j\neq0$. By Parseval and Eq.~\eqref{si:eq-inverse-symbol}, this trial energy exceeds $(4\pi^2)^{-1}\sum_{\nu\neq0}|\varphi_\kappa(\nu_1)\varphi_\kappa(\nu_2)|^2|\nu|^{-2}$ by at most $\pi^2/(48\kappa^2)$. In this sum, frequencies with $\max_i|\nu_i|\le L/\kappa$ contribute at most $(2\pi)^{-1}\ln(L/\kappa)+O(1)$. On a shell with maximum coordinate $j>L/\kappa$, one factor has square at most $(L/(2\kappa j))^2$, and the shell has at most $8j$ points, so its contribution is $O((L/\kappa)^2j^{-3})$, summing to $O(1)$. Hence $A_{\mathcal C_k}\le(2\pi)^{-1}\ln(L/\kappa)+O(1)$, and $\kappa\asymp\sqrt k$ proves the first formula. Equation~\eqref{si:eq-norm-all} implies $\eta_{\mathcal C_k}=O(1/A_{\mathcal C_k})$, and Theorem~\ref{si:thm-marked} with $k/N\to0$ gives Eq.~\eqref{si:eq-compact-bound}.
\end{proof}

\begin{theorem}[Worst case with known $k$]
\label{si:thm-worst}
On the $L\times L$ torus with known $1\le k<N$ and $N/k\to\infty$, finding any of exactly $k$ marked vertices with a fixed target success probability $0<p<1$ takes
\begin{equation}
 \Theta\Bigl(\sqrt{N\ln(N/k)}\Bigr)
\end{equation}
local steps in the worst case over the placement of the marked vertices. The same holds for deciding between no marked vertex and exactly $k$ marked vertices, the problem $\mathrm{OR}^{(k)}$ of Ref.~\cite{Aaronson2005Quantum}, with a fixed error $0<\varepsilon<1/2$. The upper bound includes $O(L)$ local steps for preparation from a known starting vertex and uses the usual membership query, whereas the lower bound grants arbitrary workspace and free queries.
\end{theorem}
\begin{proof}
\emph{Lower bound.} An algorithm that succeeds with probability $p$ on every placement does so on average over the translates of $\mathcal C_k$, to which Proposition~\ref{si:prop-compact} applies. Corollary~\ref{si:cor-marked-detection} with $\eta_{\mathcal C_k}\to0$ gives the decision bound.

\emph{Thinning.} We follow the random thinning of Ref.~\cite{Aaronson2005Quantum} (Sec.~5.4 and Lemma~5.15). Set $\beta=\lceil\sqrt{2k}\rceil$ and $w=\lfloor L/\beta\rfloor$, and partition each axis into $w$ integer intervals whose lengths differ by at most one. As $N/k\to\infty$, $w\to\infty$, and each of the $w^2$ rectangular cells has area $v_i$ between $2k$ and $C_1k$ and side length $O(\sqrt k)$, for an absolute constant $C_1$. Choose one vertex in each cell independently and uniformly, independently of the input. If cell $i$ contains $k_i$ marked vertices, its representative is marked with probability $\theta_i=k_i/v_i$. Since $\sum_ik_i=k$, the sum $\mu=\sum_i\theta_i$ satisfies $1/C_1\le\mu\le1/2$. The probability that exactly one representative is marked is
\begin{equation}
 \sum_i\theta_i\prod_{i'\neq i}(1-\theta_{i'})\ge\sum_i\theta_i\Bigl(1-\sum_{i'\neq i}\theta_{i'}\Bigr)\ge\mu-\mu^2\ge\frac1{2C_1},
\end{equation}
uniformly over all placements of the $k$ marked vertices.

\emph{Search on the thinned lattice.} The representatives form a $w\times w$ coarse torus, and adjacent representatives are joined by a fixed path of length $D=O(\sqrt k)$ inside their two cells. The coin and control operations of Tulsi's walk are on site, and its flip-flop shift is implemented by Lemma~\ref{si:lem-routing}. The membership query is applied only between completed simulated steps, when all amplitude sits on the representatives, so unselected marked vertices do not affect the simulated walk. Conditioned on exactly one marked representative, this is Tulsi's single-mark walk on the coarse torus, which succeeds with constant probability after $O(w\sqrt{\ln w})$ coarse steps, that is, after $O(Dw\sqrt{\ln w})=O(L\sqrt{\ln(N/k)})$ local steps. Since the sampling is independent of the input, the unconditional success probability is a positive constant. A membership query verifies the output, and a fixed number of independent repetitions of the entire thinning and search procedure, with fresh representatives on each trial, reaches any fixed success probability below one. Movement and preparation cost $O(L)$ per trial. Accepting only a verified marked vertex gives a one-sided decision algorithm for $\mathrm{OR}^{(k)}$.

\emph{Preparation.} The uniform state on the representatives is prepared row by row and then column by column, as in Section~\ref{si:routing}, with every coarse move routed by Lemma~\ref{si:lem-routing}. Along the first coarse row, an active and a stopped internal level are used. At representative $j=0,\dots,w-2$, the remaining active amplitude is split into fractions $1/\sqrt{w-j}$ in the stopped level and $\sqrt{(w-j-1)/(w-j)}$ in the active level, and the active level moves by one coarse edge. Only the frontier carries active amplitude, so after $w-1$ such layers and an on-site conversion at the last representative the row is uniform. The same construction runs down all columns in parallel, and an on-site rotation turns the stopped level into the initial coin and control state of Tulsi's walk. This costs $O(Dw)=O(L)$ local steps, as does moving from the starting vertex to the first representative.
\end{proof}

\begin{proposition}[Parallel windows]
\label{si:prop-windows}
Let $\ell\ge3$ divide $L$ and $k=(L/\ell)^2$, and tile the torus by $k$ disjoint open $\ell\times\ell$ windows. Let $f_1(\Mset)$ be the fraction of windows that contain exactly one marked vertex and $\bar f_1(\Mset)$ its average over the translates of $\Mset$. There is a local algorithm that, from a free uniform initial state, runs for $O(\sqrt{(N/k)\ln(N/k)})$ local steps and succeeds, on average over the translates, with probability at least $c_1\bar f_1(\Mset)$, for an absolute constant $c_1>0$. For a uniformly random $k$-subset, $\mathbb E f_1=\ell^2\binom{N-\ell^2}{k-1}/\binom Nk$, which tends to $e^{-1}$ as $k\to\infty$ and $N/k\to\infty$. For fixed $k$ its limit as $N\to\infty$ is $(1-1/k)^{k-1}$, which equals one for $k=1$.
\end{proposition}
\begin{proof}
Run the folded single-mark walk of Proposition~\ref{si:prop-fold} in every window simultaneously, without edges between windows. The uniform state gives each window weight $1/k$, every window with exactly one marked vertex succeeds with probability at least $c_1$ at the common running time, and the evolution is a direct sum over windows, so other windows cannot reduce this contribution. The probability that a given window contains exactly one member of a uniformly random $k$-subset is the hypergeometric expression, which equals $\prod_{j=0}^{k-2}[1-(\ell^2-1)/(N-1-j)]$ and has the stated limit.
\end{proof}
This is an ensemble-average statement.

\emph{Relation to known algorithms.} The dependence of particular walks on the arrangement of the marked vertices is known~\cite{Nahimovs2018Quantum,Roget2023Optimality}. For general marked sets, the grid algorithm of H{\o}yer and Komeili~\cite{Hoyer2017Efficient} is analysed through an effective hitting time and succeeds with probability $\Omega(1/\ln N)$ per run. Hitting-time frameworks~\cite{Ambainis2020Quadratic,Apers2021Unified} reach the square root of the mean hitting time with logarithmic overhead and explicit setup costs. A continuous-time construction keeps a polylogarithmic overhead when converted to constant success probability~\cite{Apers2022Quadratic}. In the present accounting, in which preparing or reflecting about a global initial state is not an on-site operation, these guarantees do not provide an $O(\sqrt{\thit(\Mset)})$ algorithm with constant success probability for every marked set.

\section{Sharpened speed limits}
\label{si:sharp}
\begin{proposition}[Commutator identity]
\label{si:prop-identity}
On a connected vertex- and edge-transitive graph with $N\ge2$ vertices and degree $d$, for every local step $U$ with on-site part $U_{\rm d}$ and every normalized $a$,
\begin{equation}
 \frac1N\sum_m\norm{[\KK_m,U]a}^2=\frac{2A}{dN}\bigl(1-\mathrm{Re}\,\langle U_{\rm d}a,Ua\rangle\bigr).
 \label{si:eq-identity}
\end{equation}
The left side is at most $4A/(dN)$. If the graph has no triangles, $\mathrm{Re}\,\langle U_{\rm d}a,Ua\rangle=\norm{U_{\rm d}a}^2$, and the upper bound improves to $2A/(dN)$.
\end{proposition}
\begin{proof}
Let $C_{xy}=\sum_mK(x,m)K(y,m)=[(\Lap^+)^2]_{xy}$. Vertex transitivity gives $C_{xx}=S^2$, and edge transitivity a common value $C_e$ whenever $x\sim y$. Since $\Lap C=\Lap^+$, its diagonal gives $d(S^2-C_e)=A$. Expanding the square,
\begin{equation}
 \sum_m\norm{[\KK_m,U]a}^2=\sum_x\sum_{y,y'}(S^2-C_{xy}-C_{xy'}+C_{yy'})\,\langle U_{xy}a_y,U_{xy'}a_{y'}\rangle .
\end{equation}
The first term contributes $S^2\norm{Ua}^2=S^2$, and the last contributes $\sum_{y,y'}C_{yy'}\langle a_y,(U^\dagger U)_{yy'}a_{y'}\rangle=S^2$. Locality gives $\sum_yC_{xy}U_{xy}a_y=C_e(Ua)_x+(S^2-C_e)U_{xx}a_x$, so the two middle terms sum to $-2C_e-2(S^2-C_e)\mathrm{Re}\,\langle U_{\rm d}a,Ua\rangle$. The total is $2A/d$ times the bracket in Eq.~\eqref{si:eq-identity}. Since $\norm{U_{\rm d}}\le1$, the bracket is at most two.

If the graph has no triangles, write $U_{\rm o}=U-U_{\rm d}$. The block $(U_{\rm o}^\dagger U_{\rm o})_{xy}=\sum_zU_{zx}^\dagger U_{zy}$ with $z\neq x,y$ requires a common neighbour $z$ of $x$ and $y$, so it vanishes whenever $x\sim y$. Since the blocks of $U_{\rm d}^\dagger U_{\rm d}$ and $I$ vanish for $x\neq y$, the blocks of $U^\dagger U=I$ at neighbouring $x,y$ give $(U_{\rm d}^\dagger U_{\rm o}+U_{\rm o}^\dagger U_{\rm d})_{xy}=0$. This operator has no other nonzero blocks, hence it vanishes, and $\mathrm{Re}\,\langle U_{\rm d}a,U_{\rm o}a\rangle=0$.
\end{proof}

\emph{Hitting time.} Combining Eq.~\eqref{si:eq-identity} with the endpoint estimate of Theorem~\ref{si:thm-bound} gives
\[
 T\ge\frac12\bigl(1-\sqrt{1-p}-N^{-1/2}-\eta\bigr)\sqrt{\thit},
\]
with $1/2$ replaced by $1/\sqrt2$ when the graph has no triangles. For fixed $p>0$ and $\eta\to0$, this implies $T=\Omega(\sqrt{\thit})$ without a degree factor. The endpoint condition is still needed, because vertex and edge transitivity alone do not force $\eta\to0$, as the cycle shows. On the complete graph, where $A=(N-1)/N^2$ and $\eta=1/\sqrt{N-1}$, this bound recovers $\Omega(\sqrt N)$.

\emph{Square torus.} Here $d=4$, and Eq.~\eqref{si:eq-identity} has coefficient $A/2N$. For $L\ge4$ the torus has no triangles, giving $|\Delta F|\le\sqrt{A/2N}$ and
\begin{equation}
 T\ \ge\ \Tmin=\frac{\bigl(1-\sqrt{1-p}-L^{-1}\bigr)LA-S}{\sqrt{A/2}},\qquad \Tmin=\bigl(1-\sqrt{1-p}\bigr)\sqrt{\frac{N\ln N}{2\pi}}-O\Bigl(\sqrt{\frac N{\ln N}}\Bigr) .
 \label{si:eq-sharp-bound}
\end{equation}
The commutator bound $A/2N$ is attained whenever $U_{\rm d}a=0$, for instance for a pure shift, so sharper control of the commutator alone cannot improve the sharpened speed limit, although the constant in Eq.~\eqref{si:eq-sharp-bound} may still be improvable by other means. For $L=3$ the general bound is $A/N$, and $A/2N$ can be exceeded. For example, the local step $U$ that acts as $\tfrac23J_3-I_3$ on one row and as the identity elsewhere, with $J_3$ the all-ones matrix, maps the uniform row state $a$ to itself while $U_{\rm d}a=-a/3$, giving $2A/3N$. The equal edge correlations used in the proof need not hold on graphs that are only vertex-transitive, or for marked patterns without the axis-exchange symmetry up to translation discussed in Section~\ref{si:marked}.

\section{Continuous-time evolution}
\label{si:continuous}
Let the unmarked run evolve under a common self-adjoint nearest-neighbour Hamiltonian $H_t$, and the marked run under $H_t+V_{m,t}$, where $V_{m,t}$ is self-adjoint and supported on $\HH_m$. Write $H_t^{\rm hop}=H_t-\bigoplus_xH_{xx,t}$ for the hopping part and assume $\norm{H_t^{\rm hop}}\le J(t)$. For well-defined normalized evolutions, for example bounded and piecewise-continuous Hamiltonians,
\begin{equation}
 \dot F=\frac{\mathrm i}N\sum_m\bra{a}[H_t,\BB_m]\ket{b^m}-\frac{\mathrm i}N\sum_m\bra{a}\BB_mV_{m,t}\ket{b^m}.
\end{equation}
The second term vanishes since $\BB_mV_{m,t}=0$, and $\mathrm i[H_t,\BB_m]$ is Hermitian. Since $\KK_m$ is scalar on each $\HH_x$, $[\KK_m,H_t]=[\KK_m,H_t^{\rm hop}]$. As the block rows of $H_t^{\rm hop}$ have norm at most $J(t)$, the proof of Lemma~\ref{si:lem-speed} yields $N^{-1}\sum_m\norm{[\KK_m,H_t]a}^2\le2J(t)^2A/N$, and the local-step part of the proof of Theorem~\ref{si:thm-bound} gives $|\dot F|\le J(t)\sqrt{2A/N}$. Integration proves Eq.~\eqref{si:eq-bound} with $T$ replaced by $\int J(t)\,dt$. For a constant hopping budget $J$ the evolution time is at least the right side of Eq.~\eqref{si:eq-bound} divided by $J$. The same argument applies to Theorem~\ref{si:thm-marked} with oracle terms supported on the marked set.

\emph{Comparison with controlled intermittent quantum walks.} The search algorithm of Ref.~\cite{Li2026Universal} illustrates the distinction between queries and motion. Theorem and equation numbers in this paragraph are those of that reference. The algorithm uses $O(\epsilon^{-1/2})$ queries and a total evolution time $O(\lambda_2^{-1}\epsilon^{-1/2})$, where $\epsilon$ is a lower bound on the marked fraction (Theorems~1 and~4 and Eq.~(25)). On the square torus $\lambda_2=4\sin^2(\pi/L)=\Theta(N^{-1})$, so with $\epsilon=k/N$ this guarantee on the evolution time is $O(N\sqrt{N/k})$. The approximate reflection about the uniform state costs an evolution time $O(\lambda_2^{-1}\log(1/\delta))$ at accuracy $\delta$ (Eq.~(21)), and this time counts as motion. In the accounting of this section, the general controlled evolution $\sum_{r=0}^{2^a-1}\ket r\bra r\otimes e^{\mathrm i r\Lap t}$ has hopping norm $(2^a-1)\norm{\mathrm{Adj}}$, which enters $\int J(t)\,dt$. The search construction uses one control qubit per such evolution, with further controls given by projections of norm one, so on the square torus its hopping norm is at most four. Ancillary rotations and control gates act on site and leave $F$ unchanged. The construction therefore lies within the continuous-time model above, and its guarantee $O(N\sqrt{N/k})$ is consistent with the lower bound.

\section{Numerical illustrations}
\label{si:numerics}
The numerical values quoted in the main text and in this Supplemental Material illustrate the theorems and enter no proof. Tables~\ref{si:tab-constants} and~\ref{si:tab-tulsi} and the rows of Table~\ref{si:tab-marked} for periodic subgrids and compact clusters follow from the definitions given in Section~\ref{si:lattice} and below, and the entries for random sets are statistics of the stated numbers of samples. The code, seeds and realized coordinates are included in the repository identified in the main-text Data availability statement.

\emph{Tulsi's walk.} The internal workspace $\HH_x$ at each vertex of the $L\times L$ torus is $\mathbb C^4\otimes\mathbb C^2$, for four flip-flop coin directions and a control qubit~\cite{Tulsi2008Faster}. With $\cos\delta=(1+8A)^{-1/2}$, $\ket{s}$ the uniform coin state and $\ket{\delta}=(-\sin\delta\ket0+\cos\delta\ket1)\otimes\ket s$, one iteration applies the query $I-2\ket{m,\delta}\bra{m,\delta}$ and then the local step $\mathrm{diag}(-I,W C_{\rm G})$ in the control basis. Here $C_{\rm G}$ is the Grover coin and $W$ the flip-flop shift. The query followed by the local step is the controlled iteration of Eqs.~(14)--(18) of Ref.~\cite{Tulsi2008Faster}. Since $A=(\ln N)/(4\pi)+O(1)$, our choice has $\cos\delta=\Theta((\ln N)^{-1/2})$, which gives constant success probability in $O(\sqrt{N\ln N})$ iterations by that reference. The walk starts in the uniform state of the control branch $\ket1$, which the unmarked run leaves invariant. By translation invariance, $F_t=\bra{a}\BB_0\ket{b_t^0}$. The initial state and all operators in this simulation are real in the position, coin and control basis, so $F_t$ is real. The success probability $p_t$ is the probability of finding the walker at $m$, summed over both control branches. The peak step $T$ maximizes $p_t$ over the first $\lceil1.5\pi L\sqrt A\,\rceil$ steps, $1.5$ times the two-level estimate $\pi L\sqrt A$ of the first peak.
Table~\ref{si:tab-tulsi} lists the results. For $L\le256$, the values of $T$ and $p$ agree with an independent second simulation of the same walk. At every step of every run, queries changed $F$ by less than $10^{-12}$ and the norm stayed within $10^{-9}$ of one. By definition of $\Tmin$, the ratio $T/\Tmin$ equals the drop of Table~\ref{si:tab-tulsi} divided by the mean rate. In these runs $F$ decreases monotonically up to $T$, so the mean rate is also the average of $|\Delta F|$ per local step.

\begin{table}[htbp]
\caption{Tulsi's walk on the $L\times L$ torus: peak step $T$, success probability $p$ and weighted overlap $F_T/A$ at the peak. Max rate is the largest $|\Delta F|$ of a local step up to $T$ and mean rate is the mean drop $(A-F_T)/T$ per local step, both in units of the sharpened speed limit $\sqrt{A/2N}$. Drop is the decrease $A-F_T$ divided by the decrease $A(1-\sqrt{1-p}-L^{-1})-S/L$ that the bound requires.}\label{si:tab-tulsi}
\centering\small
\begin{tabular}{@{}rrrrrrrrr@{}}
\toprule
$L$ & $T$ & $p$ & $F_T/A$ & max rate & mean rate & drop & $\Tmin$ & $T/\Tmin$\\ \midrule
16 & 32 & 0.6219 & 0.125 & 0.693 & 0.433 & 4.56 & 3.04 & 10.53\\
32 & 76 & 0.5864 & 0.031 & 0.714 & 0.447 & 4.38 & 7.75 & 9.80\\
64 & 179 & 0.5805 & $-0.092$ & 0.713 & 0.466 & 4.39 & 18.99 & 9.42\\
128 & 369 & 0.5770 & $-0.022$ & 0.708 & 0.454 & 3.84 & 43.62 & 8.46\\
256 & 809 & 0.5745 & $-0.067$ & 0.704 & 0.461 & 3.85 & 96.77 & 8.36\\
512 & 1626 & 0.5684 & 0.015 & 0.702 & 0.448 & 3.50 & 207.93 & 7.82\\
1024 & 3330 & 0.5582 & 0.057 & 0.702 & 0.440 & 3.37 & 435.75 & 7.64\\
\bottomrule
\end{tabular}
\end{table}

\emph{Marked sets.} The equilibrium potentials of $261$ marked sets were computed from Eq.~\eqref{si:eq-equilibrium-matrix} with fast Fourier transforms and a Cholesky factorization, for $L=64$, $128$, $256$, $512$ and $1024$ and $k=1$, $4$, $16$, $64$ and $256$. With $\ell=L/\sqrt k$, the sets are the periodic subgrid of spacing $\ell$, the compact $\sqrt k\times\sqrt k$ square cluster, sets with one uniformly random vertex in each of the $k$ windows of side $\ell$ of Proposition~\ref{si:prop-windows}, and uniformly random $k$-subsets. The single marked vertex $k=1$ was evaluated once per size. Each larger $k$ has eight uniformly random sets and four sets with one random vertex per window for $L\le512$, and four and two at $L=1024$. All sets passed checks of charge normalization and positivity, the constant value on $\Mset$, zero mean, Eq.~\eqref{si:eq-equilibrium}, the energy identity and the maximum principle, and periodic subgrids also matched Proposition~\ref{si:prop-periodic}. A direct solution of the Dirichlet problem of Lemma~\ref{si:lem-equilibrium} on the $8\times8$ torus reproduced $A_\Mset$ and the hitting times of Eq.~\eqref{si:eq-hit-marked} to rounding error. Table~\ref{si:tab-marked} lists the $1024\times1024$ torus with $k=64$. For uniformly random sets with $k=64$, the median of $4\pi kA_\Mset/\ln(N/k)$ falls from $1.695$ at $L=64$ to $1.327$ at $L=1024$, compatible with convergence to the periodic-subgrid value, for which this ratio tends to one. For these sets, Eq.~\eqref{si:eq-marked-bound} at $p=2/3$ is positive for $0/8$, $4/8$, $7/8$, $8/8$ and $4/4$ samples at $L=64$, $128$, $256$, $512$ and $1024$, so the endpoint terms $\sqrt{k/N}+\eta_\Mset$ are not negligible at these sizes.

\begin{table}[htbp]
\caption{Equilibrium quantities of marked sets on the $1024\times1024$ torus with $k=64$. Random rows are componentwise medians over two samples (one random vertex per window) and four samples (uniformly random). The fraction $\bar f_1$ is the translation average of the share of the $64$ windows of Proposition~\ref{si:prop-windows} that contain exactly one marked vertex. The last column lists the lower bound of Eq.~\eqref{si:eq-marked-bound} on the number of local steps at $p=2/3$, evaluated per sample and truncated at zero. The periodic subgrid also admits the sharpened bound $55.61$.}\label{si:tab-marked}
\centering\small
\begin{tabular}{@{}lrrrr@{}}
\toprule
Marked set & $A_\Mset$ & $\eta_\Mset$ & $\bar f_1$ & bound at $p=2/3$\\ \midrule
Periodic subgrid & 0.012828 & 0.0758 & 1 & 27.80\\
One random vertex per window & 0.013976 & 0.1415 & 0.50979 & 23.40\\
Uniformly random & 0.016017 & 0.2628 & 0.37380 & 13.97\\
Compact $8\times8$ cluster & 0.666146 & 0.0933 & $3.8\times10^{-6}$ & 190.04\\
\bottomrule
\end{tabular}
\end{table}

\emph{Consistency checks.} The identities and inequalities of Sections~\ref{si:potential}, \ref{si:bound}, \ref{si:routing} and~\ref{si:sharp} were also evaluated numerically. On $28$ small vertex-transitive graphs (cycles, complete graphs, hypercubes, circulants, a prism that is not edge-transitive, and rectangular and square tori), Eqs.~\eqref{si:eq-poisson}--\eqref{si:eq-edge-sum}, $S^2\le A/\lambda_2$ and the hitting times of Proposition~\ref{si:prop-meaning} hold to rounding error, below $10^{-12}$. Equation~\eqref{si:eq-identity} was checked on the $21$ graphs that are also edge-transitive, including complex local steps with on-site blocks. The stronger bound holds on the $15$ without triangles, and reflections within triangles exceed it on the other six. On tori with $L=5$, $8$, $12$ and $16$, runs with time-dependent, position-dependent local steps and queries preserve the norm and leave $F$ unchanged at queries, and obey the speed limit at every local step. The folded routing of Proposition~\ref{si:prop-fold} is bijective and local for $L=3$ to $32$.

\bibliography{references}